\documentclass[prd, aps, nofootinbib, preprintnumbers, showpacs,
superscriptaddress,twocolumn]{revtex4-1}

\usepackage{graphics,graphicx,rotating}
\usepackage[usenames]{color}
\usepackage[percent]{overpic}
\usepackage{mathrsfs,amsmath,amssymb}
\usepackage{wasysym}
\usepackage{times}
\usepackage{mathptmx}

\newcommand{\be}{\begin{equation}}
\newcommand{\ee}{\end{equation}}
\newcommand{\beq}{\begin{equation}}
\newcommand{\eeq}{\end{equation}}
\newcommand{\ber}{\begin{eqnarray}}
\newcommand{\eer}{\end{eqnarray}}
\newcommand{\bea}{\begin{eqnarray}}
\newcommand{\eea}{\end{eqnarray}}

\newcommand{\AEI}{\affiliation{Max-Planck-Institut f\"ur
    Gravitationsphysik, Am M\"uhlenberg 1, 14475 Potsdam, Germany}}

\newcommand{\jena}{\affiliation{Theoretical Physics Institute, University of Jena, 07743 Jena, Germany}}

\newcommand{\palma}{\affiliation{Departament de F\'isica, 
  Universitat de les Illes Balears, Cra.\ Valldemossa Km.\ 7.5, Palma 
de Mallorca, E-07122 Spain}}

\newcommand{\vienna}{\affiliation{Faculty of Physics, University of Vienna, Boltzmanngasse 5, A-1090 Vienna, Austria}}

\newcommand{\cardiff}{\affiliation{School of Physics and Astronomy, Cardiff University, Cardiff, CF24 3AA, United Kingdom}}

\begin{document}

\title{Simulations of black-hole binaries with unequal masses or non-precessing
spins: \\ accuracy, physical properties, and comparison with post-Newtonian results}

\author{Mark Hannam}        \vienna \cardiff
\author{Sascha Husa}        \palma
\author{Frank Ohme}         \AEI
\author{Doreen M\"uller}    \jena
\author{Bernd Br\"ugmann}   \jena

\date{\today}

\begin{abstract}
We present gravitational waveforms for the last orbits and merger of black-hole-binary (BBH) systems 
along two branches of the BBH parameter space: equal-mass binaries with equal non-precessing
spins, and nonspinning unequal-mass binaries. The waveforms are calculated from numerical 
solutions of Einstein's equations for black-hole binaries that complete between six and ten orbits before 
merger. Along the equal-mass spinning branch, the spin parameter of each BH is 
$\chi_i = S_i/M_i^2 \in [-0.85,0.85]$, and along the unequal-mass branch the mass ratio is 
$q =M_2/M_1 \in [1,4]$.
We discuss the construction of low-eccentricity puncture initial data for these cases, the properties of 
the final merged BH, and compare the last 8-10 GW cycles up to $M\omega = 0.1$ with the phase 
and amplitude predicted by standard post-Newtonian (PN) approximants. As in previous studies, we find 
that the phase from the 3.5PN TaylorT4 approximant is most accurate for nonspinning binaries. For 
equal-mass spinning binaries the 3.5PN TaylorT1 approximant (including spin terms up to 
only 2.5PN order) gives the most robust performance, but it is possible to treat TaylorT4 in such a way 
that it gives the best accuracy for spins $\chi_i > -0.75$. When high-order amplitude corrections are included, the PN 
amplitude of the $(\ell=2,m=\pm2)$ modes is larger than the NR amplitude by between 2-4\%. 
\end{abstract}


\maketitle

\section{Introduction}

One of the most urgent goals of numerical relativity is to produce simulations that 
will aid the detection of gravitational waves (GWs) from black-hole-binary mergers. The
current first generation of ground-based interferometric GW detectors is about to be 
upgraded, and the second-generation Advanced LIGO and Virgo detectors are 
expected to come online around 2014 \cite{Abadie:2010cfa,AdLIGO,AdVirgoDesign,Flaminio05}. 
Once operational, current 
event-rate calculations predict that they may observe multiple GW signals 
in one month of design-sensitivity operation \cite{Abadie:2010cfa}. Some of these signals will be 
from the inspiral and merger of two
black holes, and to find them in the detector data GW astronomers will use
matched filtering techniques, for which they require large collections of 
accurate theoretical waveforms (templates) of the physical signal. 

The GW signal from the last orbits and merger of black-hole-binary systems can only 
be calculated in full general relativity using numerical solutions of Einstein's equations. 
Since such simulations became possible in 
2005~\cite{Pretorius:2005gq,Campanelli:2005dd,Baker:2005vv}, they have
been used to explore larger regions of the black-hole-binary parameter space 
(which is parametrized by the mass ratio of the binary, the spin vector of each 
black hole, and the binary's eccentricity), with increasing levels of 
accuracy and covering increasing numbers of GW cycles before 
merger~\cite{Hannam:2009rd,Hinder:2010vn}. 
In addition to use in producing analytic waveform models for the construction of GW
search template banks, which we will discuss further in this paper, NR waveforms have 
also been useful in GW detection efforts as part of the NINJA project to test a battery of 
current GW search pipelines~\cite{Aylott:2009tn,Aylott:2009ya}.

In this paper we present simulations that cover between six and ten binary orbits before 
merger of configurations in two important sub-families of the binary parameter space:
unequal-mass binaries in which the black holes are not spinning, and equal-mass 
binaries where the black holes have equal spins either aligned or anti-aligned with the 
binary's orbital angular momentum. 

Following a brief summary of numerical methods in Sec.~\ref{sec:methods}, in 
Sec.~\ref{sec:eccentricity} we extend the method we developed in~\cite{Husa:2007rh}
to produce low-eccentricity parameters for spinning binaries. This 
method is based on integrating the PN equations of motion from a separation
where quasi-circular parameters are sufficiently accurate, up to the binary separation where
we wish to begin a full numerical simulation, 
and then using the momenta at that separation from the PN integration as the initial momenta
of the full numerical simulation. We now incorporate the highest-order known 
spin contributions, but find that these are still not accurate enough, and develop
a method to further refine the PN predictions. This allows us to produce simulations
of unequal-mass nonspinning and equal-mass spinning binaries with 
eccentricities of $e \apprle 0.004$. 
 
In the Samurai study~\cite{Hannam:2009hh} it was
shown that current numerical simulations for the equal-mass nonspinning case
are well within the accuracy requirements for detection with ground-based 
experiments. That study also showed that the agreement of numerical results
between different codes was consistent with the error estimates of each code 
--- and so a complete error analysis of a set of numerical simulations can 
confidently be considered as providing the uncertainty in those waveforms with 
respect to the true physical waveforms. In Sec.~\ref{sec:accuracy} we study the errors in 
our unequal-mass and equal-mass-spinning waveforms, and conclude that 
these waveforms are also well within the accuracy requirements for GW detection.

We also estimate the phase accuracy of our simulations, using a number of different methods.
The phase error accumulates quickly during the inspiral and even faster during the merger,
and small errors at a given separation or frequency are amplified during the further
evolution. On the other hand, the absolute value of the GW phase is not directly observable,
and phases from different simulations can be aligned in different ways, e.g. between
two suitably chosen fixed frequencies during the evolution. Correspondingly, the estimated
phase errors 
show a dramatic dependence on such alignment effects, and it is therefore useful to
phrase error estimates in a number of different ways.
An example relevant to a comparison with PN results
would be the time-domain phase error over a given set of GW cycles,
while for GW detection we might be more interested in the mismatch-error
in the waveform at a given binary mass with respect to a given detector. 

Having described the production of our numerical waveforms, and established their
accuracy, we summarize in Sec.~\ref{sec:properties} the physical properties
of the configurations we have studied: the mass and spin of the final merged 
black hole, the recoil (in the unequal-mass cases), and the radiated energy
and its distribution among the dominant and sub-dominant harmonics. 

One of the motivations for producing black-hole-binary simulations is to use
the resulting waveforms for the construction of analytic waveform models. 
All such models are based in some way on the expectation that PN 
approximations will be sufficiently accurate up until a few orbits before merger,
and so PN (or effective-one-body) results can be used to model the early 
inspiral, and numerical results can be used to calibrate a model for the merger.
Such a procedure first requires a measurement of the accuracy of 
PN results as the binary approaches merger. We do this in 
Sec.~\ref{sec:pncomparison}, where we compare the phase and amplitude of our
NR results with the corresponding PN predictions over the 8-10 cycles
prior to the point where the GW frequency reaches $M\omega = 0.1$, about
1.5 orbits before merger. This extends our previous studies of the equal-mass 
nonspinning binary~\cite{Hannam:2007ik,Gopakumar:2007vh} and equal-mass 
binaries in the orbital-hangup configuration~\cite{Hannam:2007wf}.

\section{Numerical methods} 
\label{sec:methods} 

We performed numerical simulations with the BAM code
\cite{Brugmann:2008zz,Husa:2007hp}. 
The code starts with black-hole-binary puncture initial data 
\cite{Brandt:1997tf,Bowen:1980yu} generated using a pseudo-spectral 
elliptic solver~\cite{Ansorg:2004ds}, and evolves them with the $\chi$-variant of the
moving-puncture \cite{Campanelli:2005dd,Baker:2005vv} version of the BSSN
\cite{Shibata:1995we,Baumgarte:1998te} formulation of the 3+1 Einstein 
evolution equations. Spatial finite-difference derivatives are
sixth-order accurate in the bulk \cite{Husa:2007hp}, Kreiss-Oliger
dissipation terms converge at fifth order, and a fourth-order Runge-Kutta
algorithm is used for time evolution. 
The gravitational waves emitted by the binary are calculated from the
Newman-Penrose scalar $\Psi_4$, and the details of our implementation of
this procedure are given in \cite{Brugmann:2008zz}.

In each simulation, the black-hole punctures are initially a coordinate distance
$D$ apart, and are placed on the $y$-axis at $y_1 = -qD/(1+q)$ and $y_2 = D/(1+q)$,
where $q = M_2/M_1$ is the ratio of the black hole masses in the binary, and we always choose 
$M_1 < M_2$. The masses $M_i$ are estimated from the Arnowitt-Deser-Misner
(ADM) mass at each puncture, according to the method described 
in~\cite{Brandt:1997tf}; we discuss this estimate of the black-hole masses and its
subtleties for spinning black holes in more
detail in Appendix~\ref{app:masses}. The Bowen-York punctures are given 
momenta $p_x = \mp p_t$ tangential to their separation vector, and $p_y = \pm p_r$
towards each other. The latter momentum component accounts for the (initially small) radial 
motion of the black holes as they spiral together. One essential question
in setting up our simulations is the determination of the parameters $(p_t,p_r)$
that lead to non-eccentric quasi-circular inspiral. We will discuss our procedure
to generate low-eccentricity parameters in Sec.~\ref{sec:eccentricity}. 

The grid setup is similar to what we have used in~\cite{Brugmann:2008zz}, and using
the notation introduced there, the simulations discussed in this paper all use a
configuration of the form
$\chi_{M\eta=2}[l_1\times N:l_2\times 2N:6]$.  This indicates that the 
simulation used the $\chi$ variant of the moving-puncture method, $l_1$ nested
mesh-refinement boxes with a base value of $N^3$ points surround each black hole,
and $l_2$ nested boxes with $(2N)^3$ points surround the entire system, and there are
six mesh-refinement buffer points. The $\eta$ parameter in the BSSN system is $M\eta = 2$.
The choices of $N$, $l_1$, $l_2$ and the resolutions
are given in Tab.~\ref{tab:grids}. The resolution around the puncture is denoted by 
$M_1/h_{min}$, which is the resolution with respect to the \emph{smallest black hole}, 
$M_1$. The puncture of the second black hole will have the same numerical resolution, 
but if the black hole is bigger, $M_2>M_1$, then it will effectively be better resolved. 

The one exception to this setup is a second convergence series for mass ratio $q=4$ 
(see the last row in Tab.~\ref{tab:grids}). These simulations use a grid configuration in 
which the effective finest resolution is the same for both black holes. This is  achieved by 
putting different numbers of refinement boxes around each puncture. As $M_1$
is four times smaller than $M_2$, we use two more boxes (the resolution doubles from 
box to box) around the smaller black hole than we do for the larger one.

Far from the sources, the meaningful length scale is
the total mass of the binary, $M = M_1+M_2$, and so the resolution on the coarsest level is
given by $h_{max}/M$. 

In our previous study of $\chi_i > 0$ cases~\cite{Hannam:2007wf} we found that extra resolution
was required around the punctures when the black holes have high spin, $|\chi_i| \apprge 0.75$.
In the newer $\chi_i < 0$ simulations we use high resolution around the puncture in all cases. 
Note also  that in only two cases ($q=2,3$), is the outer boundary causally disconnected from the 
physical system for the entire length of the simulation. This can be seen by comparing the time when
the GW signal reaches its peak amplitude in Tab.~\ref{tab:configurations}, and the location of the 
outer boundary in Tab.~\ref{tab:grids}.

\begin{table*}
\caption{\label{tab:grids}
Summary of grid setup for numerical simulations. The grid parameters follow the
notation introduced in \cite{Brugmann:2008zz}; see text. $M_1/h_{min}$ denotes the resolution on the 
finest level with respect to the \emph{smallest black hole}, while $h_{max}/M$ is the resolution on the 
coarsest level with respect to the \emph{total mass}, $M = M_1+M_2$. The outer boundary of the 
computational domain is at $x_{i,max}/M$, where $x_i = \{x,y,z\}$. In general $l_1$ indicates the 
number of moving refinement levels around each puncture, and $l_2$ the number of large refinement
levels that encompass both punctures. The one exception is the second $q=4$ series, which uses
three refinement levels around the puncture of the large black hole, and five around the other.
}
\begin{ruledtabular}
\begin{tabular}{|| l | c | c | c | c | c ||}
Configuration & $N$ & $(l_1,l_2)$ & $M_1/h_{min}$  & $h_{max}/M$ & $x_{i,max}/M$  \\[10pt]
\hline
\multicolumn{6}{||l||}{Equal-mass simulations}\\
\hline
$\chi_i = -0.85$ & 72,80,88 & (6,5) & $48,53.3,58.67$ &  $10.67,9.6,8.73$ & $774$ \\
$\chi_i = -0.75$ & 80,88,96 & (6,5) & $53.3,58.67,64$ &  $9.6,8.73,8.0$ & $774$ \\
$\chi_i = -0.50$ & 64,72,80 & (6,5) & $42.67,48,53.3$ &  $12.0,10.67,9.6$ & $774$ \\
$\chi_i = -0.25$ & 64,72,80 & (6,5) & $42.67,48,53.3$ &  $12.0,10.67,9.6$ & $774$ \\
$\chi_i = 0$ & 64,72,80 & (5,5) & $21.3,24,26.7$ &  $12.0,10.67,9.6$ & $774$ \\
$\chi_i = 0.25$ & 80,88,96 & (5,5) & $26.7,29.33,32$ &  $9.6,8.73,8.0$ & $774$ \\
$\chi_i = 0.50$ & 80,88,96 & (5,5) & $26.7,29.33,32$ &  $9.6,8.73,8.0$ & $774$ \\
$\chi_i = 0.75$ & 64,72,80 & (6,5) & $42.67,48,53.3$ &  $12.0,10.67,9.6$ & $774$ \\
$\chi_i = 0.85$ & 64,72,80 & (6,5) & $42.67,48,53.3$ &  $12.0,10.67,9.6$ & $774$ \\
\hline
\multicolumn{6}{|| l ||}{Unequal-mass simulations }\\
\hline
$q=2$ & 70,80,88 & (5,7)   & $23.3,26.67,29.33$ &  $29.26,25.6,23.27$  & $2063$ \\
$q=3$ & 70,80,88 & (5,7)   & $23.3,26.67,29.33$ &  $21.94,19.2,17.45$  & $1547$ \\
$q=4$(a) & 70,80,88 & (5,7)   & $23.3,26.67,29.33$ &  $17.55,15.36,13.96$ & $1237$ \\
$q=4$(b) & 80,88,96 & (3/5,7) & $26.67,29.33,32.0$ &  $15.36,13.96,12.8$  & $1237$      
\end{tabular}
\end{ruledtabular}
\end{table*}

\section{Specification of low-eccentricity initial parameters}
\label{sec:eccentricity}

Puncture initial data typically consist of the analytic Bowen-York solution
to the momentum constraint~\cite{Bowen:1980yu} (which allows for the construction
of multiple boosted, spinning black holes), and a numerical solution of the 
Hamiltonian constraint in puncture form~\cite{Brandt:1997tf}. To produce the data
for the simulations that we discuss in this paper we used the single-domain
spectral elliptic solver described in~\cite{Ansorg:2004ds}. In this approach the 
black holes' angular and linear momenta may be directly specified, while their
masses are specified indirectly. Some subtle issues related to the estimation of 
black-hole masses are discussed in Appendix~\ref{app:masses}. 

We wish to simulate black holes following non-eccentric inspiral. Eccentricity
cannot be so easily defined in full general relativity as in Newtonian theory, since
radiation reaction precludes the existence of circular orbits, and gauge effects
mean that any definition based on the coordinate motion of the black holes 
(or the punctures) will not be unique. Having said that, all definitions of eccentricity
based on the gauge-invariant gravitational-wave signal (see \cite{Mroue:2010re} for a 
thorough discussion of the choices available) should agree on zero-eccentricity 
inspiral, and we have found in previous work~\cite{Husa:2007rh} that definitions based 
on the coordinate motion are acceptable at the level of accuracy that we are 
interested in, which is eccentricities on the order of $e \sim 10^{-3}$. For this 
work we estimate the eccentricity from the orbital frequency of
the puncture motion by the maximum of
$e_\omega = (\omega(t) - \omega_c(t))/(2 \omega_c(t))$, where $\omega_c(t)$ is
an estimate of the non-eccentric frequency based on a curve 
fit through the numerical data~\cite{Husa:2007rh}. 

To produce low-eccentricity inspiral we need to know the appropriate initial momenta to 
give the black holes. In puncture simulations
the most effective way to do this seems to be to estimate the momenta
using PN theory. The simplest approach is to consider only a 
conservative PN model (i.e., without radiation reaction, and therefore
without inspiral), and to calculate the momenta consistent with circular orbits
at a given coordinate separation. We will refer to these as \emph{quasi-circular
(QC)} parameters. We have used QC parameters in the 
past~\cite{Brugmann:2008zz,Gonzalez:2006md,Brugmann:2007zj} for binaries no more
than three orbits away from merger, at which point small eccentricities are hard
to detect. For binaries that undergo five or more orbits before merger, it becomes
clear that the QC momenta result in noticeable eccentricities. 

One way to improve the parameters is to use a PN approach 
that includes the effects of radiation reaction. A straightforward method to do this
is by time integration of the PN equations of motion. One
begins with QC parameters  for a binary with a large separation ($D \geq 30M$), 
and integrates the PN equations 
of motion for two point particles until they have reached the separation at which we
wish to start a full numerical simulation. At that point we read off
the parameters from the PN calculation and use those in our black-hole evolution
code. We will refer to such parameters as \emph{PN inspiral (PN)} parameters. 
In~\cite{Husa:2007rh} we demonstrated that, in the equal-mass nonspinning case,
PN inspiral parameters using a 3PN accurate Hamiltonian
\cite{Jaranowski:1997ky,Damour:2001bu,Damour:2000kk} (see also 
\cite{Blanchet:2000ub,deAndrade:2000gf,Blanchet:2002mb}) and 3.5PN 
accurate radiation flux \cite{Blanchet:1997jj,Blanchet:2001aw,Blanchet:2004ek}
lead to inspirals with eccentricities 
$e \sim 0.002$. Similar results were also obtained for unequal-mass nonspinning binaries up to
mass ratio 
$q = M_2/M_1 = 4$; these were first used in~\cite{Ajith:2007kx}, and are discussed in more detail here. 
PN inspiral parameters were also used successfully to produce low-eccentricity simulations of one 
precessing-spin configuration in~\cite{Campanelli:2008nk}.

When we extended our studies to spinning binaries 
in~\cite{Hannam:2007wf} we found that the same procedure did not work so well.
At that time the PN equations included only leading-order (LO)
contributions to the spin-orbit and spin-spin Hamiltonians
\cite{Barker1970,Barker1974,Barker1979,Kidder:1995zr,Damour:2001tu,Poisson:1997ha},
and spin-induced radiation flux terms as described in \cite{Buonanno:2005xu}, 
see also \cite{Kidder:1995zr,Poisson:1997ha}.
Note that there can be ambiguities in the literature assigning PN orders
to the spin-terms in the Hamiltonian. Following \cite{Faye:2006gx}
we assign 1.5PN order to the LO spin-orbit term in the
Hamiltonian and 2PN order to the leading-order spin-spin term.

We find that the corresponding LO PN inspiral parameters lead to inspirals with 
eccentricities $e \sim 0.009$ for spins
parallel to the orbital angular momentum (and $\chi_i = S_i/M_i^2 = 0.5$), and 
$e \sim 0.03$ for the corresponding case with spins anti-parallel to the 
orbital angular momentum. In fact, in many cases  
much lower eccentricities were achieved with the supposedly cruder QC 
parameters, and these were used for the final results in~\cite{Hannam:2007wf};
in spinning cases the QC parameters include only leading-order spin
effects~\cite{Kidder:1995zr,Brugmann:2007zj}, and we have
not examined the effects of using NLO QC parameters. 
For the anti-parallel-spin cases, both the QC and LO PN inspiral parameters
lead to eccentricities that were too high to be seriously considered as 
``quasi-circular inspiral''.

For this work we have incorporated recent results
\cite{Blanchet:2006gy,Faye:2006gx,Damour:2007nc}, and include 
next-to-leading order (NLO) spin terms in the PN equations of motion.
We have also included the flux contribution due to the energy flowing in to 
the black holes, which appears at the relative 2.5PN order, as derived in 
Ref.~\cite{Alvi:2001mx}.

The improvement when using NLO PN inspiral parameters is dramatic, as shown in
Table.~\ref{tab:ereduction}. We see that in the $\chi_i = -0.5$ case, the QC 
parameters lead to an eccentricity of $e \approx 0.015$. The LO PN inspiral
parameters give even higher eccentricity, $e \approx 0.03$. When NLO spin 
terms are included, however, the eccentricity drops to $e \approx 0.004$, 
i.e., a reduction by almost an order of magnitude.
\begin{table}
\caption{\label{tab:ereduction}
Choices of the initial momenta and resulting eccentricity for the equal-mass $\chi_i = -0.5$ case.
We find that the quasi-circular (QC) parameters yield an eccentricity an 
order of magnitude larger than we desire, while the leading-order (LO) PN inspiral parameters are 
twice as bad. Incorporating NLO spin effects dramatically reduces the 
eccentricity to $e\sim 0.004$, and a further iteration based on PN predictions reduces 
it by a further 25\%. 
}
\begin{tabular}{|| c | l | l | l ||}
\hline
Parameters & $p_x$ & $p_y (\times 10^{-4})$ & $e$ \\
\hline
QC                   & 0.08469         & 0                       & 0.015 \\ 
PN (LO)          & 0.08612         &  $-5.824 $       & 0.03 \\
PN (NLO)       & 0.08500         & $-5.250  $       & 0.004 \\
PN$_+$         & 0.08512        & $-5.258    $      & 0.003 \\
PN$_-$          & 0.08487        & $-5.242    $      & 0.008 \\
\hline
\end{tabular}
\end{table}

However, this eccentricity is still twice what we can achieve in the nonspinning
case. One approach to reduce further the eccentricity would be to employ an
iterative procedure like that used for excision data in~\cite{Pfeiffer:2007yz}. 
We do not attempt this procedure, for the following reasons. The excision data
considered in \cite{Pfeiffer:2007yz} are adapted to the gauge that will be used for 
their subsequent evolution, and in particular already possess the coordinate motion
consistent with their motion along a quasi-circular inspiral. Puncture initial data, 
by contrast, start out with no coordinate motion. After the simulation begins, the 
puncture {\it wormholes} evolve into puncture 
{\it trumpets}~\cite{Hannam:2006vv,Hannam:2006xw,Hannam:2008sg}, and acquire
some coordinate motion which, after roughly one orbit, corresponds to the motion 
consistent with quasi-circular inspiral~\cite{Husa:2007rh}. Measuring the orbital 
eccentricity in order to apply an iteration procedure therefore requires performing 
puncture simulations well beyond one orbit, and even then the reduction in 
eccentricity typically converges very slowly. In addition, it is only really practical to 
estimate the eccentricity using the puncture coordinate motion, which, as should 
be clear from the preceding discussion, is purely a gauge effect and  
so may not be able to be used 
to reduce the physical eccentricity to an arbitrarily low level. 
It turns out that a similar effect 
occurs with excision data, as recently reported in~\cite{Mroue:2010re}.  Note, however,
that the results in~\cite{Mroue:2010re} arise from a different gauge condition to that used in 
moving-puncture simulations, and any conclusions they draw about the correspondence 
(or lack thereof) between the coordinate and physical motions of the black holes may
not apply to moving-puncture results. We will consider this further in 
future work~\cite{Husa:2010aa}.

For these reasons, we have attempted an alternative approach, based on
extracting further information from PN theory. We will illustrate our approach using the
same case that we have discussed above. Here the black-hole spins are $|\chi_i| = 0.5$ 
and directed anti-parallel to the binary's orbital angular momentum.
The initial coordinate separation of the punctures is $D = 12.5M$. A first simulation
is performed using NLO PN inspiral parameters.  
For reference, these are, as given in 
Table~\ref{tab:ereduction}, $p_t/M = 0.0850$ for the tangential momenta, 
and $p_r/M = -5.250 \times 10^{-4}$ for the radial momenta. 

We then calculate the eccentricity of the first orbits in this simulation. As we
stated above, we find $e \approx 0.004$. To reduce the eccentricity further, we 
now return to PN theory. If we solve the PN equations of motion starting at $D=12.5M$,
using $(p_r,p_t)/M = (0.085,-5.25\times10^{-4})$, we will of course recover the 
same non-eccentric PN inspiral as when we first calculated these parameters.
We now ask the question, ``How much would these parameters have to vary, in
order to produce the eccentricity we saw in our NR simulation?'' 
We assume that 
a variation in $(p_r,p_t)$ that \emph{produces} an eccentricity of $e = 0.004$ in the 
PN integration, starting at the same separation as the NR simulation, will give us 
approximately the correct magnitude of momenta
variation to \emph{remove} the eccentricity in the NR simulation.
Common simplifying characteristics of the situations we have in mind are that the variation in 
separation due to eccentricity, during half an orbit, say, is smaller than the variation due to
the inspiral; and that the tangential momentum is much larger than the radial momentum.

What we do now is simply adjust both the radial and tangential
momenta by some factor $k$ until the \emph{PN evolution}
produces an eccentricity of $e \approx 0.004$. For this case we find that 
$k \approx 1.0015$, i.e., by 0.15\%. We then assume that this is close to the error 
in the parameters that we have used in our NR simulation,
and modify those also by 0.15\%. We perform two additional simulations, one in which the 
momenta are increased by 0.15\%, and one in which the momenta are reduced by 0.15\%. 
The two enhanced PN-parameter choices are denoted by ``PN$_{\pm}$'' 
in Table~\ref{tab:ereduction}. 

We find, in this case, that the eccentricity is reduced when the momenta were
reduced, and we achieve $e \approx 0.003$. This is very close to the 
eccentricity we achieved in the equal-mass nonspinning case, and we
consider that acceptable. We also experimented with repeating this
iteration procedure, but it did not noticeably reduce the eccentricity; at
this level it may be possible to more cleverly modify separately the 
radial and tangential momenta, but it may also turn out that further refinement
in the {\it puncture-motion} eccentricity will not improve the true physical 
eccentricity, as indicated by the GW signal. 
At some level of accuracy it will
also be necessary to adjust the radial and tangential momenta by different
factors. Another shortcoming of the method we have used here is that we do not
explicitly use phasing information when adjusting the eccentricity of the orbit, e.g.
to determine whether momenta should be increased or decreased without having to perform
two further simulations.
On reason why this is difficult, is
because in the initial gauge of the simulation the punctures are stationary on
the numerical grid, and it takes $\sim$ one orbit for their motion to asymptote to 
a trajectory consistent with the physical motion of the black holes. 
We have found the results to be acceptable for all cases we have considered here. Further work
on improving this procedure is underway, and preliminary results on an improved method that
also uses phasing information have been presented recently \cite{GR19talk}.  

The procedure we have described here was performed on all of the anti-hangup cases, and the final
parameters are given in Table~\ref{tab:configurations}. For these cases we also
indicate the eccentricities that were achieved from the raw PN-inspiral parameters.
Also shown are the parameters for the nonspinning case presented in \cite{Hannam:2007ik},
the hang-up cases described in \cite{Hannam:2007wf}, and a set of nonspinning unequal-mass 
simulations. In the unequal-mass simulations, the eccentricity was found to be sufficiently 
low with the raw PN inspiral parameters, and no further modifications were made.
This also suggests that while we expect the PN approximation to 
deteriorate for larger mass ratios, this deterioration is not large at $q=4$.

However, a second series of $q=4$ simulations was also performed, using 
EOB parameters as described in~\cite{Walther:2009ng}. Whereas PN inspiral
parameters lead to an eccentricity of $e \approx 0.0038$, the EOB parameters
lead to a lower eccentricity of only $e \approx 0.003$. This appears to be consistent 
with the  expectation that EOB methods retain their accuracy at higher mass ratios better
than PN methods, although we note that the uncertainty in the eccentricity calculation is
$5 \times 10^{-4}$, and so the two values agree within uncertainty. Results from much 
higher-mass-ratio simulations are necessary to definitively compare the performance 
of EOB and PN parameters.

\section{Summary of numerical simulations} 
\label{sec:summary}

In this section we summarize the two sets of configurations that we studied. 

The first comprised equal-mass ($q = 1$) binaries, with equal
spins directed either parallel ($\chi_i > 0$) or anti-parallel ($\chi_i < 0$) to the orbital angular
momentum of the binary. The spins considered were $|\chi_i| = \{0,0.25,0.5,0.75,0.85\}$. 
When the spins are \mbox{(anti-)}parallel to the orbital angular momentum, they will not precess, 
making such cases a relatively simple sub-family of the total black-hole-binary parameter
space. And when $q=1$, the system possesses enough symmetry that the simulation can
be performed on only a quarter of the full physical domain, $z>0, y>0$; this symmetry 
is also reflected in the fact that in these configurations the center-of-mass of the system 
does not move, and the final black hole does not experience any recoil. 

In addition, the choice of equal spins also yields an important sub-family of 
configurations: we found in~\cite{Ajith:2009bn} that it is possible to rather
 accurately model any
non-precessing binaries with \emph{unequal spins} using essentially only a 
\emph{mass-weighted sum of the spins} of the binary, motivated by numerical evidence 
from~\cite{Vaishnav:2007nm,Brugmann:2007zj,Reisswig:2009vc}, 
and PN theory~\cite{Arun:2008kb}.
Therefore for the purposes of producing waveform models for GW detection
with current ground-based detectors, it is sufficient to simulate only binaries where the black 
holes have equal spins. Recall also that the use of Bowen-York-puncture data limits us to 
black holes with spins 
$|\chi_i| \apprle 0.92$~\cite{Dain:2008ck,Lovelace:2008tw,Hannam:2009ib,Hannam:2006zt}, 
and that since errors due to the presence of junk radiation increase with higher spins, 
the maximum spin that we treat in these simulations is $|\chi_i| = 0.85$. 

The second set of configurations we consider is nonspinning binaries with unequal masses, 
$q = \{2,3,4\}$. Now the symmetry of the system is reduced, and the simulations require
half of the physical domain, $z>0$. The center-of-mass of the system can move, and does
due to the nature of the asymmetry of the GW emission from these systems, and the final
black hole ``recoils'' (or is ``kicked'') relative to the original center-of-mass of the binary. 
We will discuss further the recoil in our unequal-mass simulations in Sec.~\ref{sec:recoil}.

Of the simulations discussed in this paper, a small subset were first presented elsewhere. 
The equal-mass nonspinning simulations were described in detail in~\cite{Hannam:2007ik}, 
and the $|\chi_i| > 0$ cases in~\cite{Hannam:2007wf}. The remaining simulations have not yet been published, although they have all be used as part
of other studies. The $q=2$ simulations were used to study parameter-estimation accuracy for 
LISA~\cite{Babak:2008bu}. In addition, all of these
simulations have been used to build phenomenological waveform families.  Some of 
the nonspinning-binary data were used in~\cite{Ajith:2007kx,Ajith:2007xh} to produce
nonspinning phenomenological waveforms, and in~\cite{Damour:2008te} to calibrate an
effective-one-body (EOB) model. 
It should be noted, however, that in those works 
less-accurate simulations of the higher-mass-ratio cases were used, and in particular
the accuracy of the $q=4$ data used in~\cite{Damour:2008te} were not sufficiently accurate
to conclusively test the physical fidelity of the EOB model. 
Also, all of the waveforms presented here were used to produce the first non-precessing-spin
phenomenological model presented in~\cite{Ajith:2009bn}, and the follow-up study 
in~\cite{Santamaria:2010yb}. In fact, extra simulations for unequal-mass \emph{spinning}
 binaries were also necessary for that work, but we will not consider those here.

The methods we used to estimate the initial momenta for quasi-circular inspiral were
described in Sec.~\ref{sec:eccentricity}, and fall into three classes: quasi-circular (QC), 
PN inspiral (including next-to-leading order [NLO] spin terms when necessary), and 
enhanced PN$_{\pm}$ inspiral, as described in Sec.~\ref{sec:eccentricity}. For an
extra $q=4$ series, we also used parameters based on an EOB model. 
QC parameters were used to produce the 
older orbital hangup $\chi_i > 0$ simulations, which were originally 
presented in~\cite{Hannam:2007wf}; for these cases the eccentricity is the highest,
around $e \sim 0.006$. PN inspiral parameters were used for the 
unequal-mass simulations, where they were found to yield acceptably low eccentricities
of $e < 0.004$. Finally, PN$_{\pm}$-inspiral parameters were used for the 
anti-hangup cases $\chi_i < 0$, and for all cases the eccentricity is $e < 0.003$. 
The EOB parameters that were used for the second $q=4$ series lead to an 
eccentricity of $e \approx 0.003$.

\section{Accuracy of numerical simulations}
\label{sec:accuracy}

In this section we will estimate the errors in our numerical results. These will
give us some indication of both the physical accuracy of our waveforms, and
the applications for which they can confidently be used. 

The waveform can be decomposed into phase and amplitude functions, 
$\phi_{\ell m}(t)$ and $A_{\ell m}(t)$ respectively, for each spherical harmonic mode $(\ell,m)$. 
The individual harmonics of the Newman-Penrose scalar $\Psi_4$ can 
now be written as \begin{equation}
R_{\rm ex} \Psi_{4,\ell m}(t)  = A_{\ell m}(t)  e^{- i \phi_{\ell m}(t)},
\end{equation} where $R_{\rm ex}$ is the coordinate radius of the wave-extraction 
sphere. The frequency of the signal is given by $\omega_{\ell m} = d\phi_{\ell m}/dt$. The results
in this paper use GW signals extracted at $R_{\rm ex} = 90M$, unless otherwise
stated.

We will focus on the $(\ell=2,m=2)$ mode. The numerical error 
in the functions $A_{22}(t)$ and $\phi_{22}(t)$ is estimated 
by means of a convergence test: we perform simulations at three (or more)
numerical resolutions, and verify that differences between successive 
resolutions decrease at a rate consistent with the expected convergence
properties of the numerical code. Throughout the remainder of this paper, 
the functions $A(t)$, $\phi(t)$ and $\omega(t)$ will refer to the $(\ell=2,m=2)$
quantities unless otherwise stated. If a clear convergence rate is 
observed, then it is also possible to use Richardson extrapolation to 
remove the error term at the next order, and produce a yet more accurate
estimate of the true result, and to also calculate an uncertainty estimate. This
procedure was carried out for our equal-mass nonspinning data 
in~\cite{Hannam:2007ik}.

As described in Sec.~\ref{sec:methods}, we have based our grid configurations on
that used in our work on equal-mass nonspinning binaries~\cite{Hannam:2007ik}.
For that configuration, the defining number of points in the three convergence series
simulations (see Tab.~\ref{tab:grids}) was $N= \{64,72,80\}$. This was sufficient to achieve
reasonably clean sixth-order convergence, which had been identified 
in~\cite{Husa:2007hp} as the dominant order of finite-difference error in our code.
For configurations with higher mass ratios and non-zero spins, greater numerical
resolution is required. In many of the new simulations the number of grid-points
has therefore been increased. In some cases the extra resolution was sufficient to again 
achieve sixth-order convergence, but in others it was not. For these latter cases,
although we can be confident that the simulations are converging towards the 
continuum solution, we are unable to use Richardson extrapolation to estimate
the uncertainties, and must provide much more conservative error estimates. 
Although we could perform more simulations at yet higher resolution, we find that
even these conservative error estimates are within the error bounds required for 
this work and many GW-astronomy applications.

We will illustrate these points with two representative cases: one that shows clean
sixth-order convergence, and one that does not. We focus first on the GW phase, because
it is the phase error that dominates the mismatch calculations that are at the heart of
the matched-filtering technique used in GW searches,
and because the error in the \emph{amplitude} is dominated not by numerical
resolution but by the radius $R_{\rm ex}$ of the GW extraction; we will discuss
this further
in Sec.~\ref{sec:mismatcherrors}.

\subsection{GW phase}
\label{sec:phaseaccuracy}

\begin{figure*}[t]
\centering
\includegraphics[width=80mm]{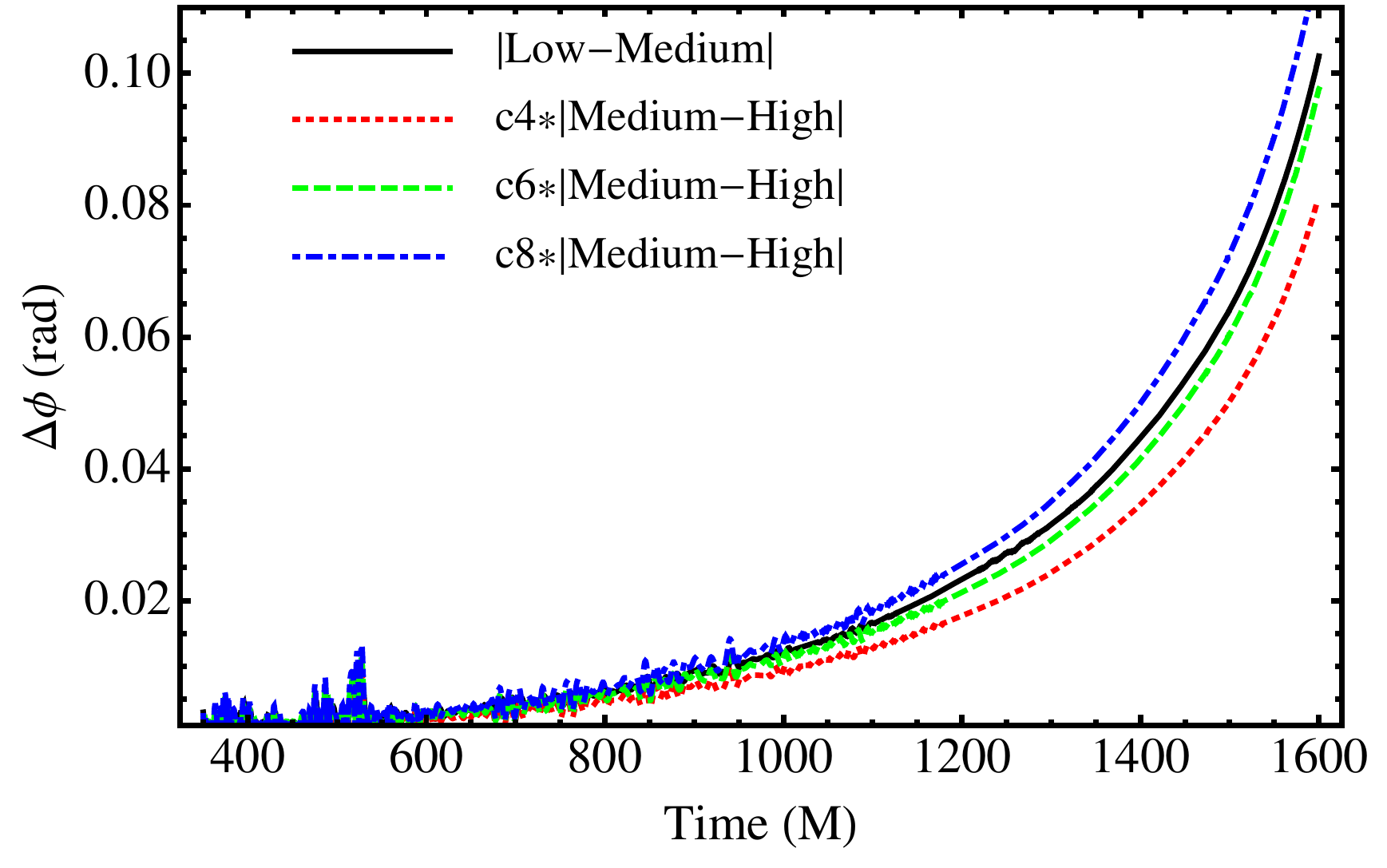}
\includegraphics[width=80mm]{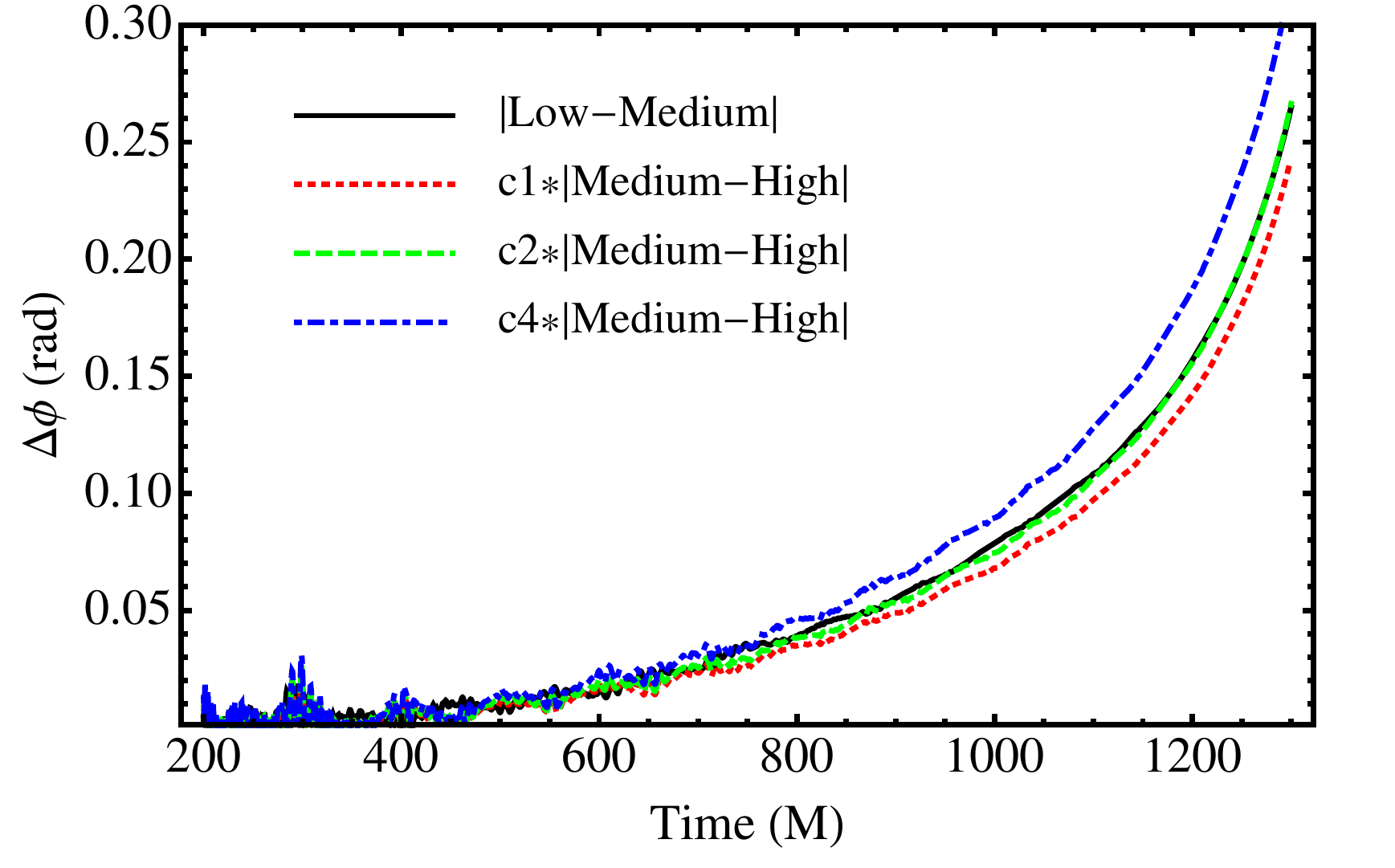}
\includegraphics[width=80mm]{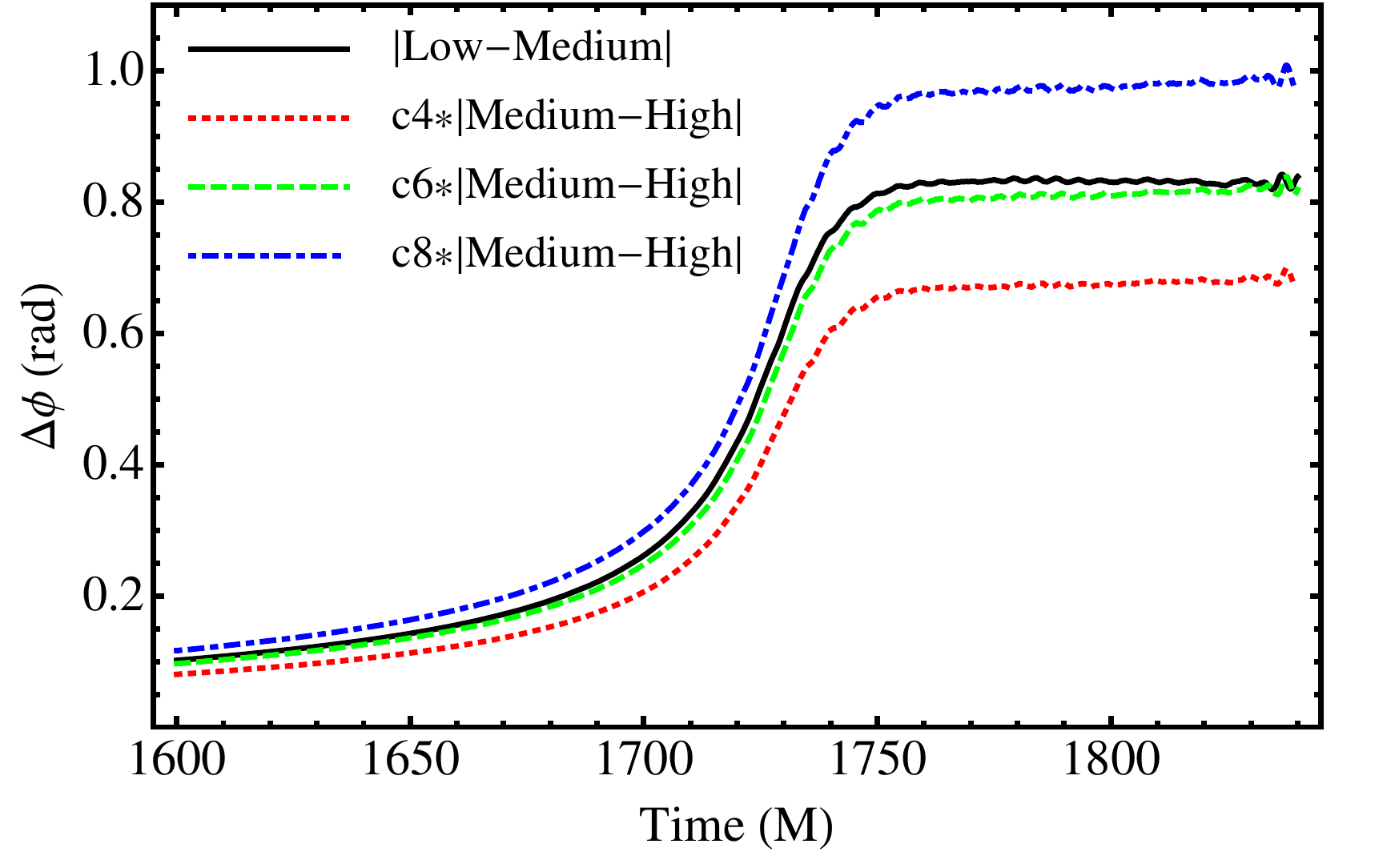}
\includegraphics[width=80mm]{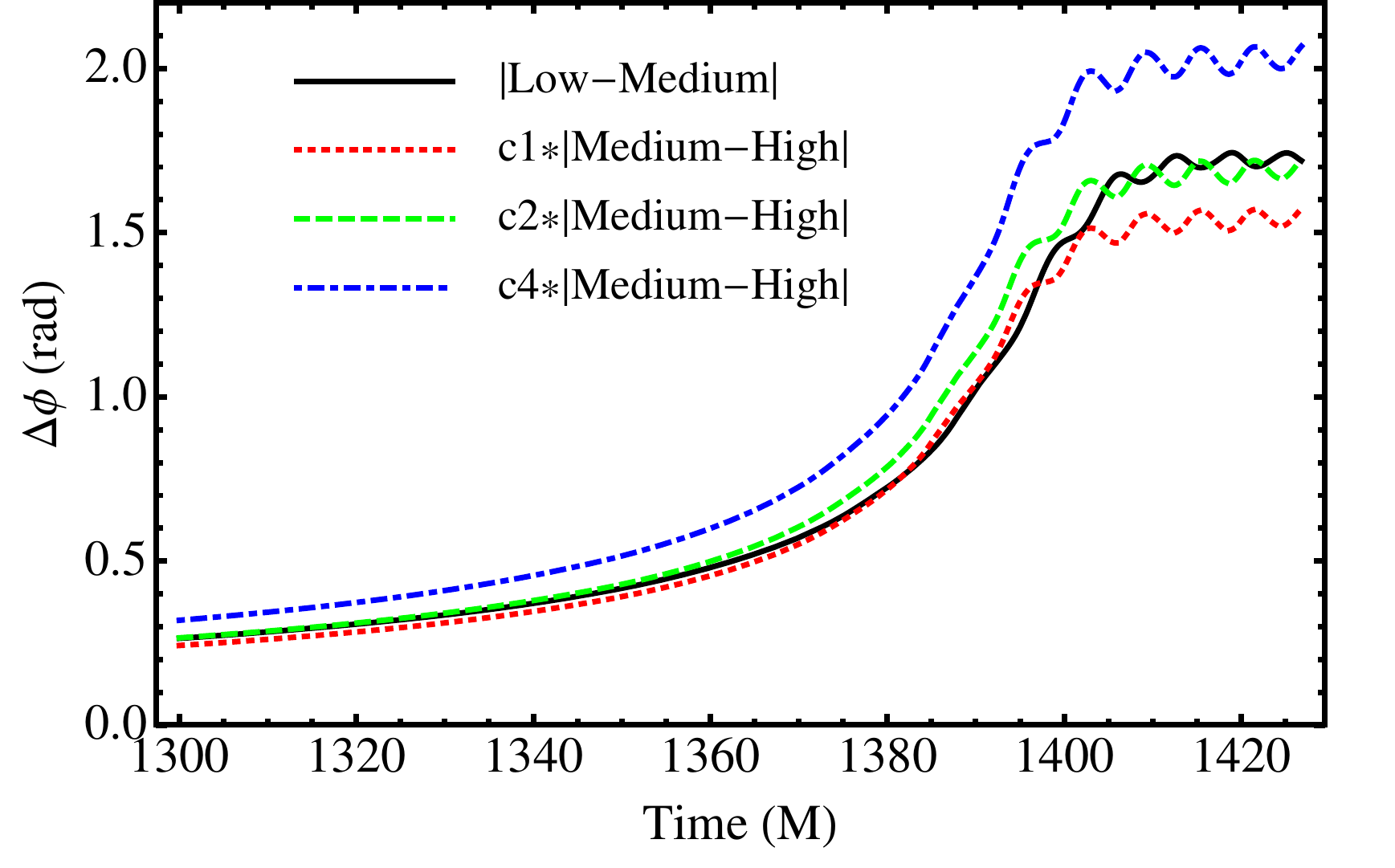}
\caption{
Convergence of the phase as a function of time for the $\chi_i = 0.5$ (left)
and $q=4$ (second convergence series) cases. The early-time behaviour is shown
in the upper plots, and the late-time behaviour in the lower plots. Scaling with 
respect to different convergence orders is shown, to illustrate how cleanly the data
exhibit a particular convergence behaviour.
In these plots $t=0$ indicates the beginning of the simulation, 
and $\Delta \phi(t=0) = 0$ in all simulations. The $\chi_i = 0.5$ case shows 
reasonably clean 6th-order convergence, and the
accumulated phase difference is $\Delta \phi = 0.43$\,rad between the medium-
and high-resolution simulations. The $q=4$ case is not yet in the 6th-order convergent
regime, and appears (erroneously) to exhibit 2nd-order convergence. The accumulated 
phase difference between the medium- and high-resolution simulations is 1.5\,rad.
}
\label{fig:cvgstd}
\end{figure*}

In studying the GW phase, we have two aims: (1) to show that our results are converging
to the continuum solution as a function of numerical resolution, and to provide 
error bounds on the GW phase, and (2) to illustrate the ambiguities inherent in estimating
the phase error. The ambiguity in estimating phase errors is already well known, but 
we discuss it further here to make the point that although the accuracy of the GW phase 
is important, any one method of estimating the ``phase error'' may tell us little about the 
waveform's accuracy for a given application.

We consider three measures of the phase error. The first is the total accumulated phase
error over the length of the simulation. This is the most natural quantity to study in a
convergence test. The second is the accumulated phase error over the ten cycles up 
to a GW frequency of $M\omega = 0.1$. We will need this error estimate to justify the 
comparison with PN results in Sec.~\ref{sec:pncomparison}, but we will also see that this
quantity is problematic when used for a convergence test, as is any realignment of the GW
phase. None of these estimates have a natural interpretation for applications in GW searches. 
There the mismatch error is the appropriate measure of the waveform's accuracy, 
and we will consider this in Sec.~\ref{sec:mismatcherrors}. 

For our convergence analysis we consider in detail two cases, $(q=1, \chi_i = 0.5)$ 
and the second $(q=4, \chi_i = 0)$ convergence series. 
These are indicative of the general features of all of the cases we
have studied. 

Standard convergence plots of the GW phase are shown in Fig.~\ref{fig:cvgstd}. 
The initial phases agree at $t=0$, and the plot shows the subsequent evolution of the phase 
disagreement between simulations at different resolutions. In the $\chi_i = 0.5$ case
(left panel), which uses relatively high resolution for a moderate spin value, 
we see reasonably clean 6th-order convergence. The $q=4$ case, however, even though
the numerical resolutions are the same, is not yet in the 6th-order-convergence regime, 
and for these choices of grid resolutions, appears to be 2nd-order convergent. We 
emphasize that this is \emph{not} a demonstration that the simulations have
entered a 2nd-order-convergence regime; while it is expected that at sufficiently high
resolutions the 2nd-order-accurate components of the code will dominate, we have not
yet performed \emph{any} simulations of any configuration with high enough resolution to 
see clean asymptotic 2nd-order convergence. All the second panel of Fig.~\ref{fig:cvgstd}
tells us is that we are not yet in the fully convergent regime, but since the results 
are converging, in the sense that the errors reduce between simulations, we can still 
make a conservative estimate of the accumulated phase error. 

In the $\chi_i = 0.5$ case, we can use Richardson extrapolation based on 6th-order
convergence, to estimate the uncertainty in the accumulated phase as 0.6\,rad. In
the $q=4$ case, however, we do not yet see 6th-order convergence. If we were to 
optimistically assume that the medium- and high-resolution simulations \emph{are}
in the convergent regime, and it is only the low-resolution simulation that is not, 
then we would estimate an accumulated phase uncertainty of 1.9\,rad based on 
Richardson extrapolation. If we instead produce a much more conservative 
error estimate based on 2nd-order Richardson extrapolation, we find 6.7\,rad. Note
that this is an order of magnitude higher than we found in the cleaner $\chi_i = 0.5$
case, and corresponds to a full GW cycle.

How are we to interpret these accumulated phase errors? They are certainly useful
in comparing simulations --- for example, the second $q=4$ convergence series is less 
accurate than the $\chi_i = 0.5$ series. But this measure of the phase accuracy
is of little additional value. In a GW application we will use only the waveform \emph{after}
the passage of the pulse of junk radiation. It is then difficult to estimate the phase error of the
resulting waveform because to do so we must first align the waveform at some point after the
beginning of the simulation. This introduces ambiguities (due to numerical noise in the 
GW phase and frequency) that may be larger than the error we ultimately want to measure;
this point is illustrated well in Sec.~V.E.2 of~\cite{Boyle:2007ft}.

When we compare with PN approximants in Sec.~\ref{sec:pncomparison}, we will be interested
in the phase error over the 8-10 GW cycles up to GW frequency $M \omega_m = 0.1$.
If we look at Fig.~\ref{fig:cvgstd}, we find that the accumulated phase difference at that frequency,
between the medium- and high-resolution simulations, is 0.1\,rad and 0.12\,rad, respectively
for the $\chi_i = 0.5$ and $q=4$ cases. However, if we instead line up the waveforms in
each convergence series at $M\omega_m = 0.1$, and measure the accumulated phase
disagreement as we go back 10 cycles, we instead find about 0.01\,rad for both configurations.
This is an order of magnitude lower than what we observe when the waveforms are aligned
at the beginning. This is an artifact of both the removal of the junk-radiation portion of the
waveform, and simply the properties of the waveform frequency functions; similar effects
are seen with different choices of alignment of PN waveforms, for which no junk radiation 
or significant numerical noise exist. 

These results demonstrate that we must be careful to choose our assessment of the phase
error consistently with the application we are interested in. For the PN phase comparison 
in Sec.~\ref{sec:pncomparison}, we compare PN and NR waveforms aligned at 
$M\omega_m = 0.1$, and so the only meaningful numerical phase error estimate that 
makes sense is that based on the same form of phase alignment. 

Noise in the numerical frequency introduces an ambiguity into the 
matching time for any phase re-alignment procedure, which makes it impossible to use the 
realigned phase as the basis of a convergence test. However, we can vary the matching 
time within its error bounds, 
measure the \emph{maximum} accumulated phase disagreement that arises from this 
process, and then use 4th-order Richardson extrapolation to provide a conservative 
error estimate in the phase. The results of this process are shown in Tab.~\ref{tab:errors}, 
and will be relevant to the analysis in Sec.~\ref{sec:pncomparison}. The same procedure
and alignment are used to give estimates of the phase uncertainty accumulated during
merger and ringdown. The table also 
shows an estimate of the total accumulated phase error, based on a convergence analysis
like that shown in Fig.~\ref{fig:cvgstd}; we repeat that this estimate has no direct relation
to any physical application, and is only useful as a means to compare the relative
accuracy of different simulations. Note that this number is \emph{not} simply the sum of the
inspiral and merger phase uncertainty estimates, and this is a clear artifact of the alignment
ambiguity in assessing phase accuracy. As such, in most physical applications, where some
realignment is implicitly performed, the effective 
total phase error may drop by an order of magnitude over the numbers shown in the table. It is
also clear that the total accumulated phase error estimates depend dramatically on whether
we see clean convergence (the one truly clean case shown in the table is $\chi_i = 0.5$;
other lower spin cases are also cleanly convergent). Nonetheless, we will see in 
Sec.~\ref{sec:mismatcherrors} that this level of accuracy is still well within the requirements
for GW detection.

\subsection{GW amplitude} 
\label{sec:FreqFns}

We now consider the GW amplitude. This plays a less important role in {\emph
detection}, but errors in the amplitude (as well as higher harmonics) will affect estimates 
of the source parameters, since all parameter errors scale with inverse signal-to-noise
ratio (SNR).

If we perform a time-domain convergence analysis of the GW amplitude, our conclusions
are biased because the apparent amplitude error is in fact a combination of both the 
amplitude \emph{and} phase errors --- if the amplitude were measured with no error by the code,
but two waveforms are out of phase, they will appear to have a non-zero amplitude error when 
compared in the time domain. 
We discussed this point in some detail in~\cite{Hannam:2007ik}, and used a parametrization
of the amplitude in terms of GW phase to reduce the effects of dephasing on the 
amplitude analysis. This works well if the phase error as a function of GW frequency is small,
but this will not always be true. We expect (from the PN and perturbation theories) that the
GW amplitude is a function of the GW frequency, and so the ideal method to measure the 
amplitude accuracy would be to reparametrize the amplitude as a function of GW frequency.

This procedure also presents problems: 
the GW frequency is a numerically noisy function during the early and late parts of the 
simulation; it is certainly not the smooth monotonically increasing function that we 
expect it to be on physical grounds. We can partially circumvent this 
difficulty by producing a smooth analytic fit of the frequency function, and considering
the GW phase and amplitude as parametrized by that function. The smoothing 
process may itself introduce numerical artifacts, and either mask or exaggerate
the convergence properties of the numerical results. But in general it is sufficient to 
allow us to calculate uncertainty estimates for our waveforms. 

Our method for modeling the GW frequency is as follows, based on an earlier version that
was used (for equal-mass, nonspinning waveforms) in the work for the Samurai 
project~\cite{Hannam:2009hh}. 
For the inspiral, we start with the analytic TaylorT3 approximant for the frequency, 
as given in~\cite{Buonanno:2006ui}. We neglect the highest-order (3.5PN) nonspinning 
term and replace it by a free parameter that will be fit to our data. In addition, 
following~\cite{Buonanno:2006ui}, we do not  specify the value of the spin, but also treat it as a free 
parameter --- remember that our goal is to produce a clean analytic fit to the frequency, and 
we are not interested in whether all of the parameters have their usual physical interpretation. 
The modified TaylorT3 frequency function is then \begin{eqnarray*}
\label{eqn:T3}
& &\Omega_{\rm PN}(\tau) \nonumber\\
& & =  \frac{1}{4} \tau^{-3/8} \left[ 
  1 
+ \left( \frac{743}{2688} + \frac{11}{32} \nu \right) \tau^{-1/4} 
- \frac{3}{10} \pi  \tau^{-3/8} \right. \nonumber  \\
& & + \left( \frac{47}{40} \frac{S}{M^2} + \frac{15}{32} \frac{\delta M}{M} \frac{\Sigma}{M^2} \right) \tau^{-3/8} \nonumber \\
& &\left.  + \left( \frac{1855099}{14450688} + \frac{56975}{258048} \nu +
  \frac{371}{2048} \nu^2 \right) \tau^{-1/2} \right. \nonumber \\
& & \left. + \left( - \frac{7729}{21504} + \frac{13}{256} \nu \right)
  \pi \tau^{-5/8}  \right. \nonumber  \\
& & + \left[ \left( \frac{101653}{32256} + \frac{733}{896} \nu \right) \frac{S}{M^2} +
\left( \frac{7453}{7168} + \frac{347}{896} \nu \right) \frac{\delta M}{M} \frac{\Sigma}{M^2} \right] \tau^{-5/8}
\nonumber \\ 
& & \left. + \left( - \frac{720817631400877}{288412611379200} + \frac{53}{200}\pi^2 
 + \frac{107}{280} \gamma \right. \right. \nonumber \\ 
& & \left. \left. - \frac{107}{2240} \ln \left(
       \frac{\tau}{256} \right)  
+ \left( \frac{25302017977}{4161798144} - \frac{451}{2048} \pi^2
\right) \nu \right. \right. \nonumber \\
& & \left. \left. - \frac{30913}{1835008} \nu^2 + \frac{235925}{1769472} \nu^3\right) \tau^{-3/4}
  + a \tau^{-7/8}   \right], \label{eqn:omfit}
\end{eqnarray*} where $\nu = m_1 m_2 / M^2$ is the symmetric mass ratio, $S = S_1 + S_2$ 
is the total spin parallel to the orbital angular momentum, $\Sigma = M(S_2/m_2 - S_1/m_1)$,
and $\delta M = m_1 - m_2$. (Note that in the Samurai paper, the PN frequency 
formula Eqn.~7 is missing an overall factor of two.) In the cases we consider here, the spins are 
nonzero only 
in the equal-mass case, and the spins are always equal to each other, so the 
$(\delta M \, \Sigma)$ terms do not contribute. The function $\tau$ is usually given 
by $\tau = \nu (t_c - t)/(5M)$, and $t_c$ is interpreted as the ``time of coalescence''
in standard PN theory, although a more appropriate term would be ``time of divergence''. 

In order to produce a formula that can be fit through our data, we redefine $\tau$ as
\begin{equation}
\tau^2 = \frac{\nu^2 (t_c - t)^2}{25 M^2} + d^2,
\end{equation} where both $t_c$ and $d$ are free parameters that are fit to the data. 
This modification of $\tau$ prevents $\Omega_{\rm PN}$ from diverging at $t=t_c$. In the
form that we have written it, $\Omega_{\rm PN}$ is now symmetric about $t=t_c$, which is
certainly not physically realistic, but beyond this point we will make a smooth transition
to a different function, which models the ringdown.

To model the ringdown phase, we modify the ansatz suggested
in \cite{Baker:2008mj}, and write the full frequency as \begin{eqnarray}
\Omega(t) & = & \Omega_{\rm PN}(\tau) + \nonumber \\ 
&& \left[\Omega_f - \Omega_{\rm PN}(\tau) \right] 
\left( \frac{1 + \tanh[\ln\sqrt{\kappa} - (t - t_0)/b]}{2} \right)^\kappa. \nonumber \\
\label{eqn:omRD}
\end{eqnarray} The constants $\{t_c,t_0,S,\kappa,a,b,\Omega_f\}$ are
parameters that are determined to produce the best fit to the
numerical data. The constant $\Omega_f$ corresponds to a fit of the
ringdown frequency, but the other parameters have no clear physical interpretation.
(Even the ``spin'' parameter $S$ really amounts to no more than a modification of the
1.5PN and 2.5PN terms in the description of the inspiral frequency.)

\begin{figure*}[t]
\centering
\includegraphics[width=80mm]{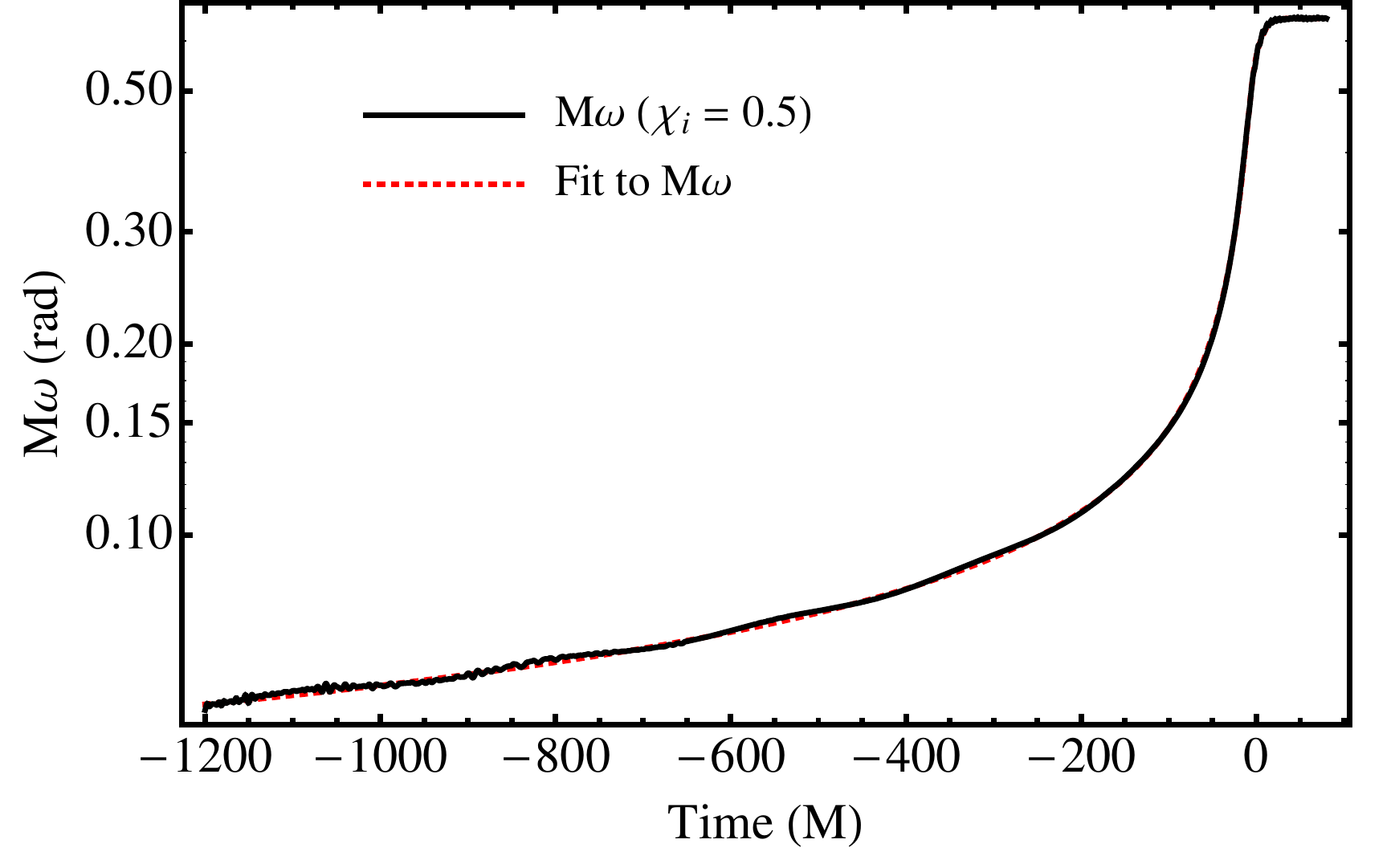}
\includegraphics[width=80mm]{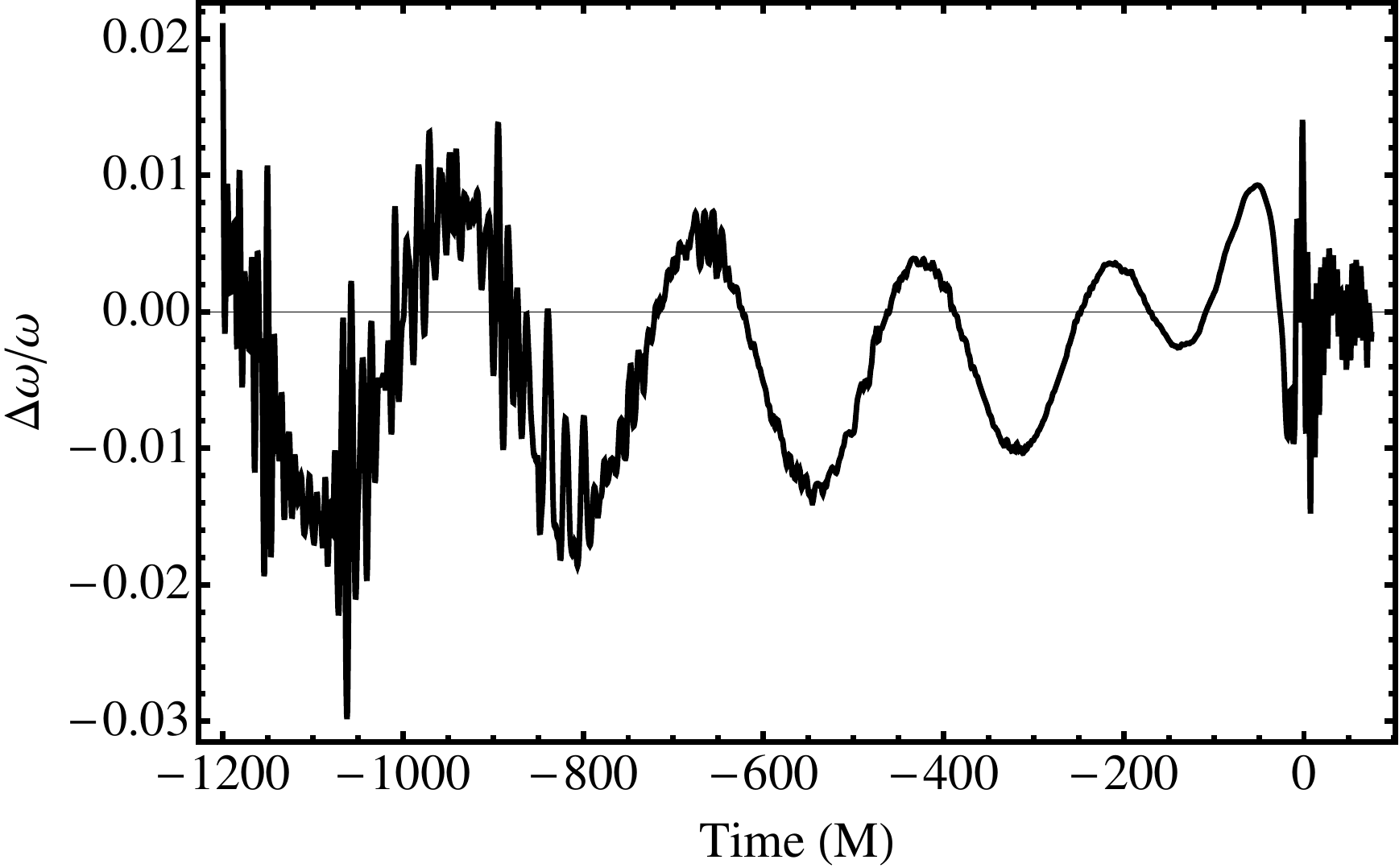}
\caption{
Analytic fit to the GW frequency for the $\chi_i = 0.5$ case. The right panel 
shows the fractional difference between the fit and numerical data. For this
configuration, the error in the fit is dominated by the residual eccentricity in 
the simulation. The dashed line indicates the point at which the amplitude 
reaches its maximum.
}
\label{fig:omfits}
\end{figure*}

Fig.~\ref{fig:omfits} shows a typical frequency fit, in this case for $\chi_i = 0.5$. We see
that the dominant error in the fit is due to the residual eccentricity in this simulation;
recall that the aligned-spin cases are based on QC parameters and have the 
highest eccentricity of all the cases we studied. The procedure does not work quite
so well in cases with high spin; the frequency evolution is not captured
so well during the early inspiral, or in the $200M$ before the peak GW
amplitude. The fitting formula (\ref{eqn:omfit}) could be modified to address this, 
and indeed the model of the transition to ringdown~(\ref{eqn:omRD}) has since been 
improved by the authors of~ \cite{Baker:2008mj,smcwcomment}. 
These issues, and the masking of eccentricity effects, mean that this frequency fit is far 
from ideal, and cannot be used for a convergence study of the amplitude. 
However, it is adequate for the purpose of providing a rough estimate of the 
amplitude uncertainty in our simulations.

\begin{figure}[t]
\centering
\includegraphics[width=80mm]{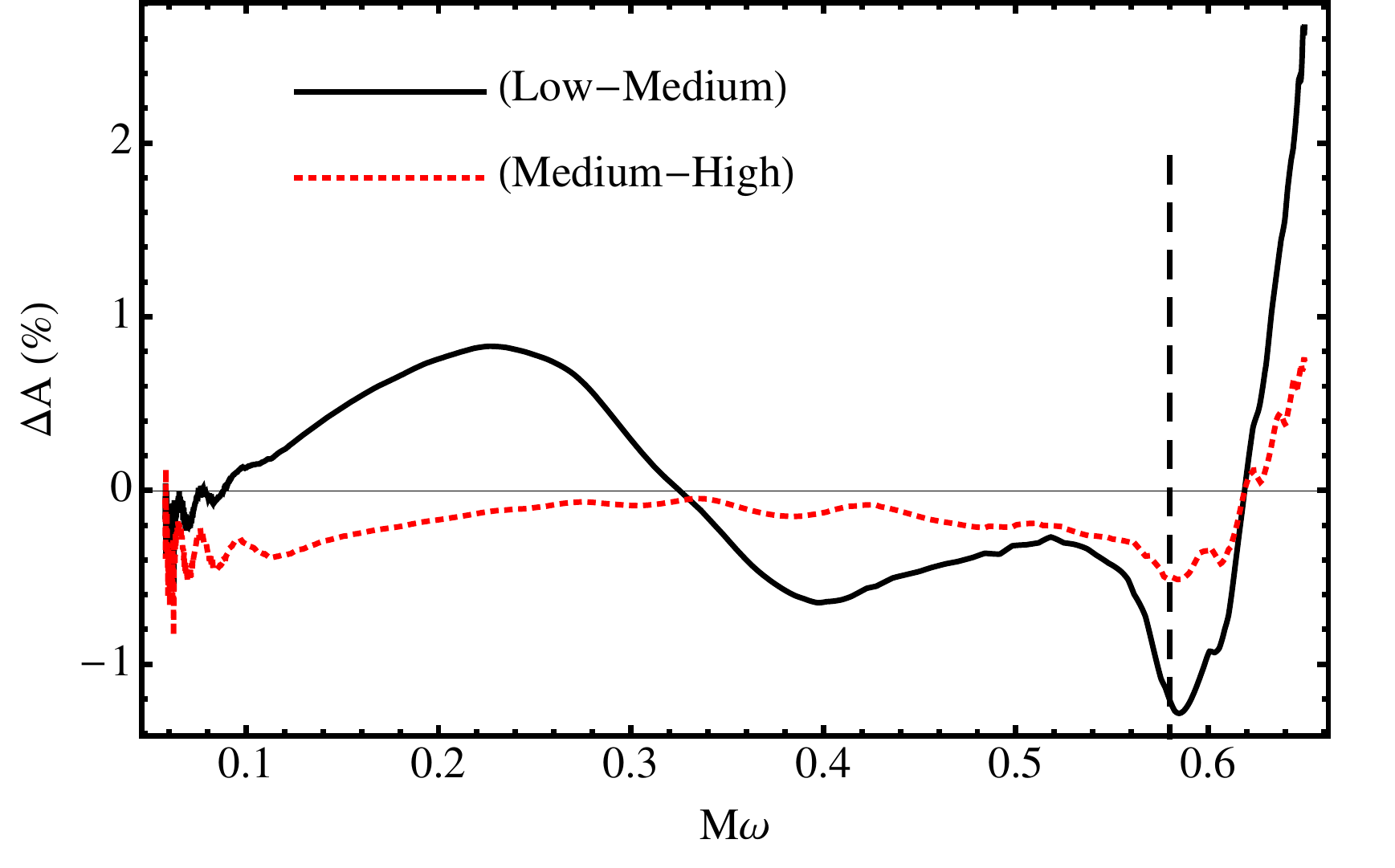}
\caption{
The amplitude error as a function of GW frequency for the $\chi_i = 0.5$ case. 
The deficiencies of the frequency fitting procedure preclude the use of $A(\omega)$ for
a convergence test, and the differences between the low-, medium- and high-resolution 
simulations are not scaled in any way. 
}
\label{fig:pp50Amp}
\end{figure}

Fig.~\ref{fig:pp50Amp} shows the differences in $A(\omega)$ with respect to resolution
for the $\chi_i = 0.5$ configuration. The figure suggests that the error in the GW
amplitude due to numerical resolution is on the order of 1\%. At late times the 
relative error grows higher, but this is beyond the frequency at which the amplitude 
reaches a maximum (indicated by the dashed line), and is well into the ringdown of the
signal. Note that if we perform an error analysis based on the time-domain amplitude,
then the maximum error between the medium- and high-resolution simulations is 
around 6\%, which suggests that it is indeed dominated by the phase error. 

Estimates for the amplitude uncertainty (using the amplitude parametrized by GW
frequency) are given in Tab.~\ref{tab:errors}. In all cases the GW signal 
was extracted at $R_{\rm ex} = 90M$; we will discuss the errors due to the use of a finite
extraction radius in the next section.

\begin{table*}
\caption{\label{tab:errors}
Estimates of uncertainty in phase and amplitude. The phase uncertainty accumulated during the inspiral is based
on an alignment of the GW phase at $M\omega = 0.1$, and includes only the ten GW cycles up to that frequency, 
for consistency with the analysis in Sec.~\ref{sec:pncomparison}. The same alignment is used for the phase 
uncertainty of the merger and ringdown regime.
The ``complete'' phase uncertainty is a conservative estimate of the total accumulated phase error over the 
entire waveform, and is \emph{only} relevant for relative comparisons of different simulations; see text in 
Sec.~\ref{sec:phaseaccuracy}. The amplitude uncertainties are described in Sec.~\ref{sec:FreqFns}, 
and the mismatch errors in Sec.~\ref{sec:mismatcherrors}.
}
\begin{tabular}{|| l |c|c|c|c|c||c ||}
\hline
Case         & \multicolumn{3}{|c|}{Phase uncertainty}          & \multicolumn{2}{|c||}{Amplitude uncertainty}  & Mismatch\\   
                   & \multicolumn{3}{|c|}{(radians)}     & \multicolumn{2}{|c||}{(percentage)}  &  ($\times 10^{-4}$)   \\
                   \hline
                   & inspiral & merger & complete                          & inspiral    & merger &    \\
\hline

$q=1$       & \multicolumn{6}{|c||}{} \\
\hline
 $\chi_i = +0.85$     & $0.1$    & 2.10 &   $10$  &  $0.25$ &  $5.0$ & 2.8 \\
 $\chi_i = +0.50$     & $0.05$  & 0.75 &  $1.0$ &  $0.5$    & $1.0$   & 1.0  \\
 $\chi_i = -0.50$      & $0.1$   & 0.80  & $10$  &  $0.5$    & $4.0$ &  0.8 \\
$\chi_i = -0.85$       & $0.1$   & 0.75 &  $15$  &  $0.5$    &  $2.0$ &  0.7 \\
\hline
$q=2$                       & $0.05$ & 0.2 &  $5.0$    &  $0.2$    & $1.0$ & 0.3  \\
$q=3$                       & $0.05$  & 0.3 & $10$ &  $0.4$    & $2.0$ & 2.7  \\
$q=4$(a)                  & $0.1$  &  1.5 & $15$  &  $0.25$  & $4.0$ &  3.2  \\
$q=4$(b)                  & $0.05$   & 0.8  & $7.0$  &  $0.25$  & 2.0 &  ---  \\ 
\hline
\end{tabular}
\end{table*}

\subsection{Mismatch with respect to numerical frequency and GW extraction radii} 
\label{sec:mismatcherrors}

Ultimately we are interested in the accuracy of our waveforms with respect to 
GW detection. The most meaningful way to do this is to calculate the 
\emph{faithfulness} between waveforms from different numerical
resolutions and different extraction radii. 

A calculation of the faithfulness is based on the overlap between two waveforms. 
The overlap is usually calculated in the frequency
domain. For two GW signals (in this analysis all quantities are with respect to the 
($\ell=2,m=2)$ mode of the signal) $\tilde{h}_1(f)$ and $\tilde{h}_2(f)$, we define an 
inner product weighted by the power spectral density of the detector noise, $S_n(f)$,
as~\cite{Cutler94},
 \beq
   \label{eq:scalar_prod}
 \langle h_1|h_2\rangle := 4 \, {\rm Re} \left[ \int_{f_{\rm
         min}}^{f_{\rm max}} 
     \frac{ \tilde h_1(f) 
     \tilde h_2^\ast (f)}{S_n(f)} \, df \right] \, .
 \eeq Our data represent $\Psi_4(t)$, not the wave strain $h(t)$, but the two are
 related by $\Psi_4 = \ddot h_+ - i \ddot h_\times$. Making two time integrations 
 is trivial when transforming to the frequency domain, and although this does not
 automatically remove the irritation of having to choose constants of integration
 (see~\cite{Reisswig:2010di} for a recent discussion of this problem) we have
 found that our ignorance of these constants does not affect mismatch 
 calculations~\cite{Hannam:2009hh}. 
 
Given the definition of the inner product $\langle h_1|h_2\rangle$, we normalize it
and maximize over phase and time offsets in the data. If the waveforms were equal,
then this quantity would be unity. This is the faithfulness of the waveform: it is a measure
of how ``far'' a theoretical waveform is from a supposedly true waveform with the same
physical parameters. We define the \emph{faithfulness mismatch} as the deviation
from unity: \beq
{\cal M} = 1 -  \max_{\tau,\Phi} \frac{\langle h_1 | h_2 \rangle }{\sqrt{
     \langle h_1|h_1\rangle \langle h_2|h_2\rangle }}.  \label{eqn:mismatch}
\eeq

Ideally the integrations over frequency are in the range $[0,\infty]$. When we
have a finite data set like our numerical waveforms, we also need to optimize with 
respect to the window of our data that we sample. This is discussed in further detail 
in~\cite{Hannam:2009hh}; the optimization with respect to phase and time offset is
trivial in the frequency domain when using only one the $(\ell=2,m=2)$ mode. 
In general more sophisticated maximisation 
procedures are required, for example the techniques described in~\cite{Damour:1997ub} 
for the quadrupole harmonic, and in~\cite{McWilliams:2010eq} for signals that include 
higher harmonics.

The faithfulness mismatch is calculated without any optimization over the 
intrinsic parameters of the binary. In a true GW search using a bank of theoretical templates,
one optimizes not only over time and phase shifts, but over all physical parameters included
in the template bank. Optimization over the physical (intrinsic) parameters gives the 
\emph{effectualness mismatch}, i.e., how well a waveform family will be able to detect GW
signals irrespective of whether the physical parameters are measured correctly. 
We have access to waveforms representing only one choice of 
intrinsic parameters of the binary, and so cannot perform this optimization (although we could 
easily optimize over the total mass of the binary). 
The physical-parameter-optimized effectualness mismatch will always be better than 
(or equal to) the faithfulness mismatch and so we can use
the faithfulness to set an upper bound on the error of the waveforms. 

The faithfulness mismatch between the simulations of different resolutions is negligible ---
it is below $10^{-6}$ for all relevant masses (down to about 100$M_\odot$) with respect
to the Advanced LIGO noise curve~\cite{LAL}
(we use the approximate analytical formula displayed in \cite{Ajith:2007kx}), where we choose a low-frequency cut-off of 20\,Hz. 

We also use the faithfulness to estimate the error due to the finite extraction radii. The
GW signal is extracted on spheres of radii $R_{\rm ex} = \{50,60,70,80,90\}M$, and we expect
the error relative to the true signal as $R_{\rm ex} \rightarrow \infty$ to fall off as $1/R_{\rm ex}$. 
This error is typically larger than that due to finite-difference errors, i.e., the finite extraction 
radius is the dominant source of error in the simulation. We estimate the mismatch error
by extrapolating to $R_{\rm ex} \rightarrow \infty$ the mismatches between our
finite-extraction-radii
data. We find that the maximum mismatch
(which is always at the lowest mass we consider, $100\,M_\odot$) is $2.8\times10^{-4}$. 
This is much 
 larger than the mismatch due to the numerical resolution errors, as we expect.
 The maximum mismatch in each case is given in Tab.~\ref{tab:errors}.
 
Such levels of accuracy are well within the requirements set out 
in~\cite{Lindblom:2008cm}, and also within the suggested accuracy for waveform
modeling within the NR-AR project~\cite{ninja-wiki}. This is also comparable or better 
than the level of mismatch
between the different equal-mass nonspinning waveforms (taken from independent codes)
that were studied in the Samurai project~\cite{Hannam:2009hh}, suggesting that these
waveforms are also of sufficient accuracy for GW detection purposes with current 
ground-based detectors. The accuracy requirements for parameter estimation, and for 
applications with future detectors, such as the space-based LISA~\cite{Shaddock:2009za} 
and third-generation ground-based Einstein Telescope~\cite{Punturo:2010zza}, may
be much higher, but have not yet been quantified for NR waveforms.

\section{Physical properties of the binary configurations} 
\label{sec:properties} 

Now that we have established the accuracy of our simulations, we can calculate some
of their physical properties. The most accurate are quantities calculated from the phase 
and amplitude of the leading harmonic, like the mass and spin of the 
final black hole. Less accurate are integrated quantities based on the leading subdominant 
harmonics, like the radiated energy in each mode. The gravitational recoil, which is not
only based on the higher harmonics, but on overlaps between some of the weaker harmonics, 
is the least accurate. We will consider first the general physical properties of the 
binary configurations. 

\subsection{General properties}

In Table~\ref{tab:configurations} we indicate the initial coordinate 
separation of the binary $D/M$, and the number of GW cycles before merger, $N_{\rm GW}$. 
The latter quantity is defined as $\Delta \Phi/(2\pi)$, where $\Delta \Phi$ is the accumulated GW
phase from $t = 200M$ (i.e., after the early burst of junk radiation) until the time when the wave's
amplitude reaches its maximum value. 

The configurations we simulated for this work clearly demonstrate the orbital hang-up and 
``anti-hangup'' effects for spins parallel or anti-parallel to the orbital angular momentum. The
orbital hang-up case was first studied in~\cite{Campanelli:2006uy}, and 
for a larger range of cases in~\cite{Hannam:2007wf}; the
largest spin considered was $\chi_i = 0.92$ in~\cite{Dain:2008ck}. 
One case of anti-parallel spins with $\chi_i = -0.438$ was considered in~\cite{Chu:2009md}. 

When the black holes are nonspinning, a binary with an initial coordinate separation of $D = 12M$ 
produces around 19 GW cycles before merger. When the black holes have spins $\chi_i = 0.25$, 
the merger is delayed in comparison with the nonspinning case, and a binary with the same initial 
separation produces 21.5 GW cycles before merger. Conversely, when the spins are $\chi_i = -0.25$,
the merger is accelerated, and the binary produces only 18.5 GW cycles before merger. These
trends continue as the spins are increased, and in order to produce comparable numbers of GW 
cycles in each simulation, the initial coordinate separation is increased for increasing anti-parallel
spins $\chi_i < 0$, and decreased for increasing parallel spins $\chi_i > 0$. For the highest-spin
cases, $|\chi_i| = 0.85$, an initial separation of $D=13M$ is required to produce 16 GW cycles in the 
anti-parallel case, and an initial separation of only $D=10M$ produces 20 GW cycles in the 
parallel case. 

\begin{figure*}[t]
\centering
\includegraphics[width=80mm]{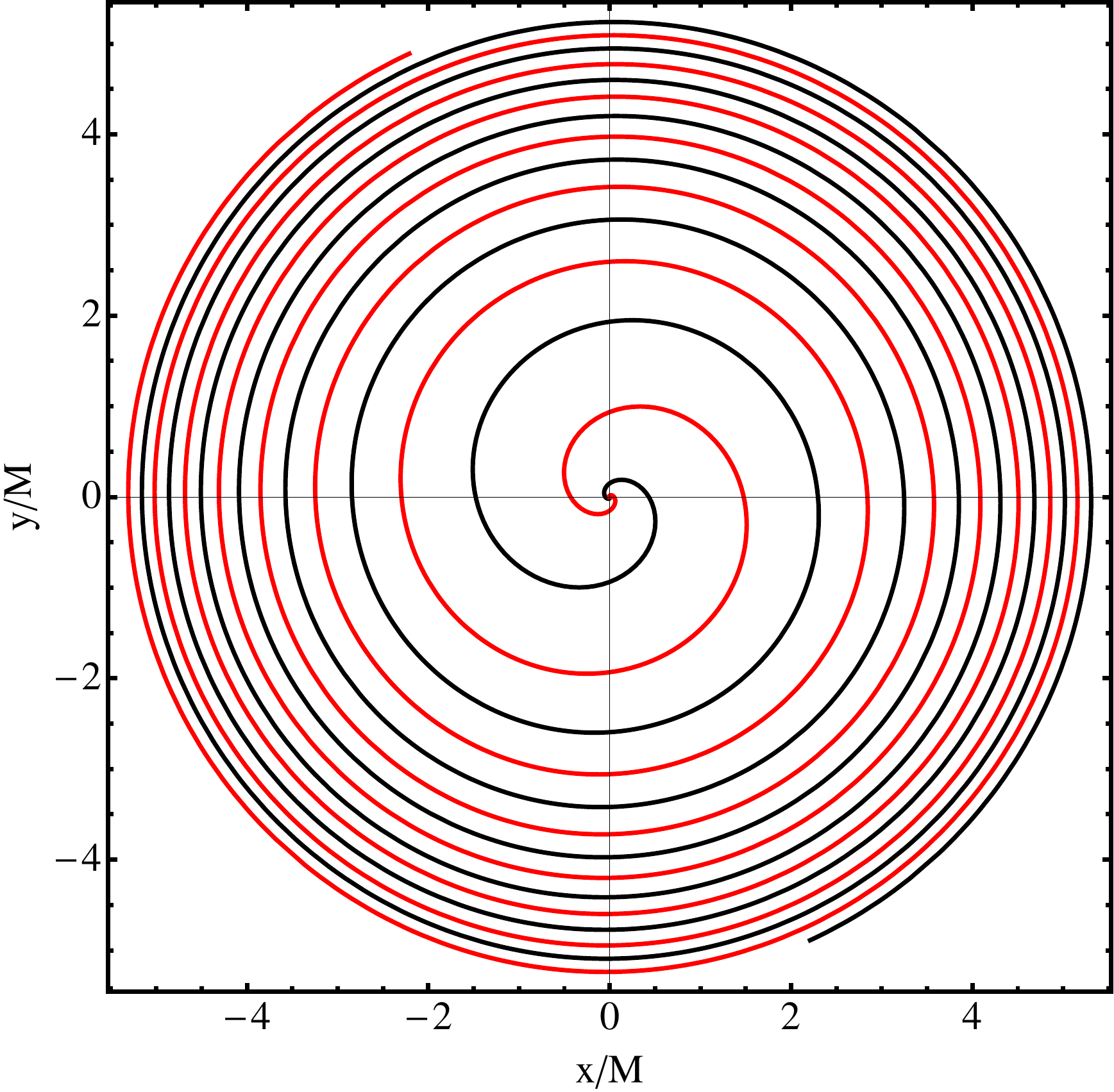}
\includegraphics[width=80mm]{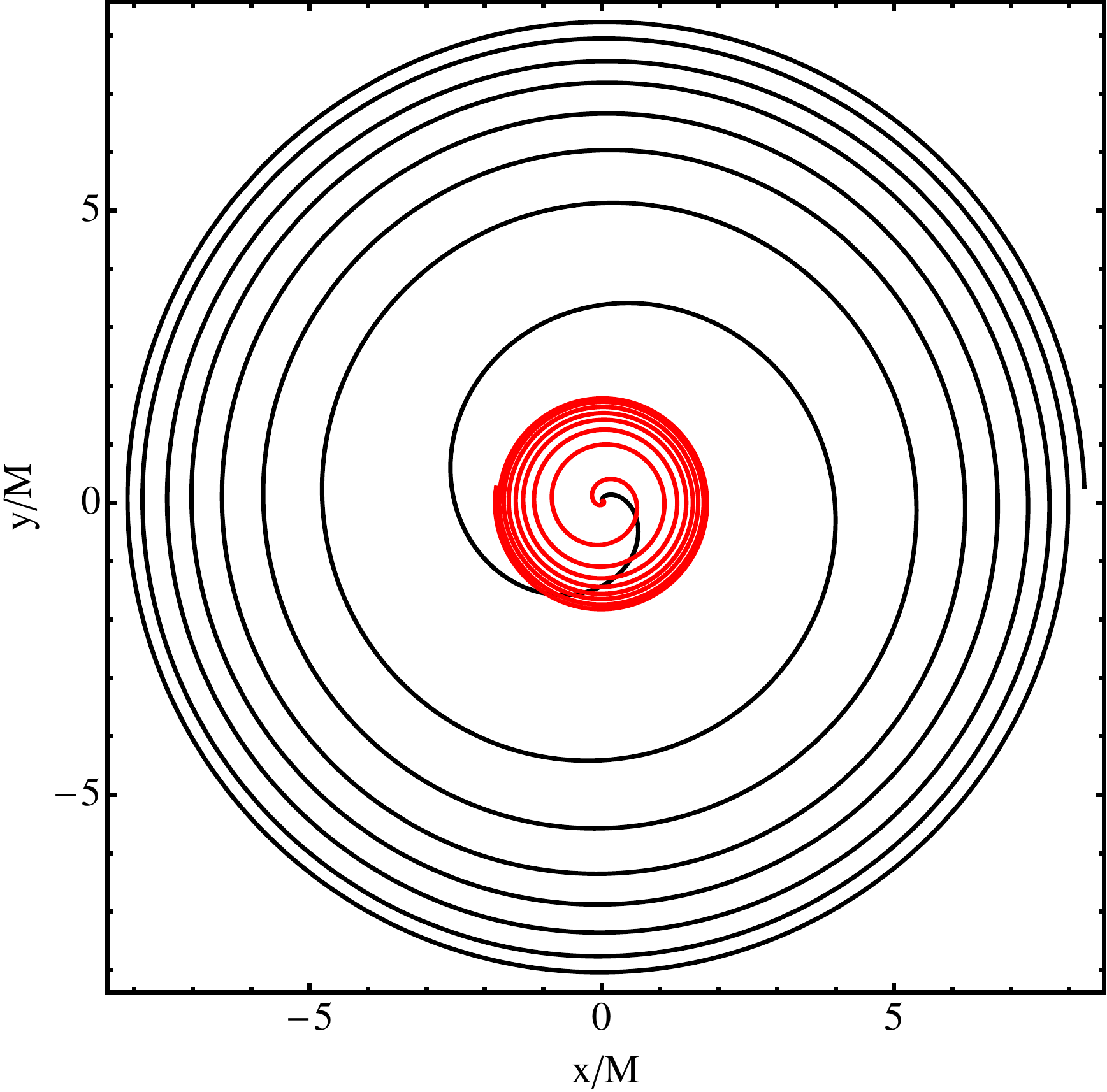}
\caption{
Puncture motion for the nonspinning binary configurations $q=1$ (left) and $q=4$
(right). The figure shows about seven orbits before merger for each system. In the 
$q=4$ case, the black line indicates the small black hole, while the red line indicates
the large black hole. Note that the transition from inspiral to plunge and merger is 
more gradual in the $q=1$ case. As the mass ratio increases, the plunge begins to 
resemble the ISCO effect that is present for extreme mass ratios.
}
\label{fig:q14tracks}
\end{figure*}

In the unequal-mass nonspinning cases, we see that the number of cycles before merger also 
varies with the mass ratio $q$. The general effect is best understood by considering the two 
extreme cases, $q=1$ and the extreme-mass-ratio case $q \rightarrow \infty$. In the 
extreme-mass-ratio case, i.e., a point particle orbiting a Schwarzschild black hole, the slow 
inspiral terminates abruptly at the innermost stable circular orbit (ISCO), and the small black 
hole plunges into the large black hole. Prior to the ISCO, the small black hole follows a slow
adiabatic inspiral, so that there are very many orbits at separations just above the ISCO, but as
soon as the small black hole passes the ISCO, there are no more orbits, only the fast plunge. 
By contrast, in the equal-mass nonspinning case, the transition from ``inspiral'' to ''plunge'' is 
very smooth, and there is no ISCO; the rate of inspiral simply increases. 
As the mass ratio is increased, the rate of the ``plunge'' increases, and the rate of inspiral prior
to merger decreases --- in other words, the dynamics approach the extreme-mass-ratio situation,
and the system gets closer to exhibiting an ISCO.  This behaviour is illustrated in 
Fig.~\ref{fig:q14tracks}. 

\subsection{Final mass and spin}
\label{sec:finalspin}

The final mass of the merged black hole can be estimated from the energy lost through 
gravitational radiation. Given the total (ADM) energy in the initial data, $E_{\rm ADM}$, 
and the radiated energy $E_{\rm rad}$, we know that the final spacetime must contain 
the energy $M_f = E_{\rm ADM} - E_{\rm rad}$. Since the final spacetime contains only a single 
stationary (i.e., non-radiating) Kerr black hole, $M_f$ must be the mass of that black hole. 

We calculate the radiated energy $E_{\rm rad}$ on each of the five extraction spheres
$R_{\rm ex} = \{50,60,70,80,90\}M$, and extrapolate the result to $R_{\rm ex} \rightarrow \infty$
assuming that the error falls of as $1/R_{\rm ex}$. This assumption is most consistent with the
data at the largest extraction radii, and so we use only $R_{\rm ex} = \{70,80,90\}M$ for the
fit, and include also $R_{\rm ex} = 60M$ to assess the robustness of the result. Once
$E_{\rm rad}$ has been estimated for each resolution, we find that the results converge
at roughly 4th-order, although since the convergence is not extremely clean, we use
\mbox{2nd-,} 4th- and 6th-order Richardson-extrapolated values to estimate the uncertainty 
in the value from the highest-resolution simulation. In all cases we consider the uncertainty
in the radiated energy to be about 2\%. The values of the mass of the final black hole are
given in Tab.~\ref{tab:configurations}.

To estimate the spin of the final black hole, we make use of analytic results that give the
quasi-normal ringdown frequency $M_f  \omega_{\rm RD}$ in terms of the black-hole spin, 
$a_f/M_f$~\cite{Berti:2005ys}. In the ringdown regime, the GW signal behaves as 
$\sim \exp(-i \omega_{\rm RD} t)$,
where $\omega_{\rm RD}$ consists of a real part (which is the frequency of the ringdown waveform),
and an imaginary part, which describes the rate of exponential fall-off. Given the final mass
$M_f$ and the ringdown waveform, we can estimate the final spin using either the
exponential decay-rate of the wave's amplitude, or the wave's frequency $M\omega$ 
in the ringdown stage. We find that matching to the ringdown frequency gives the most
accurate results, in the sense that both methods agree within uncertainties, but 
the uncertainty estimates are smaller when we match to the ringdown frequency. In general
our final spin estimates have an uncertainty of 1\%, although it is a little smaller in the $\chi_i = 0.85$ 
case, which we now consider in more detail. 

\begin{figure}[t]
\centering
\includegraphics[width=80mm]{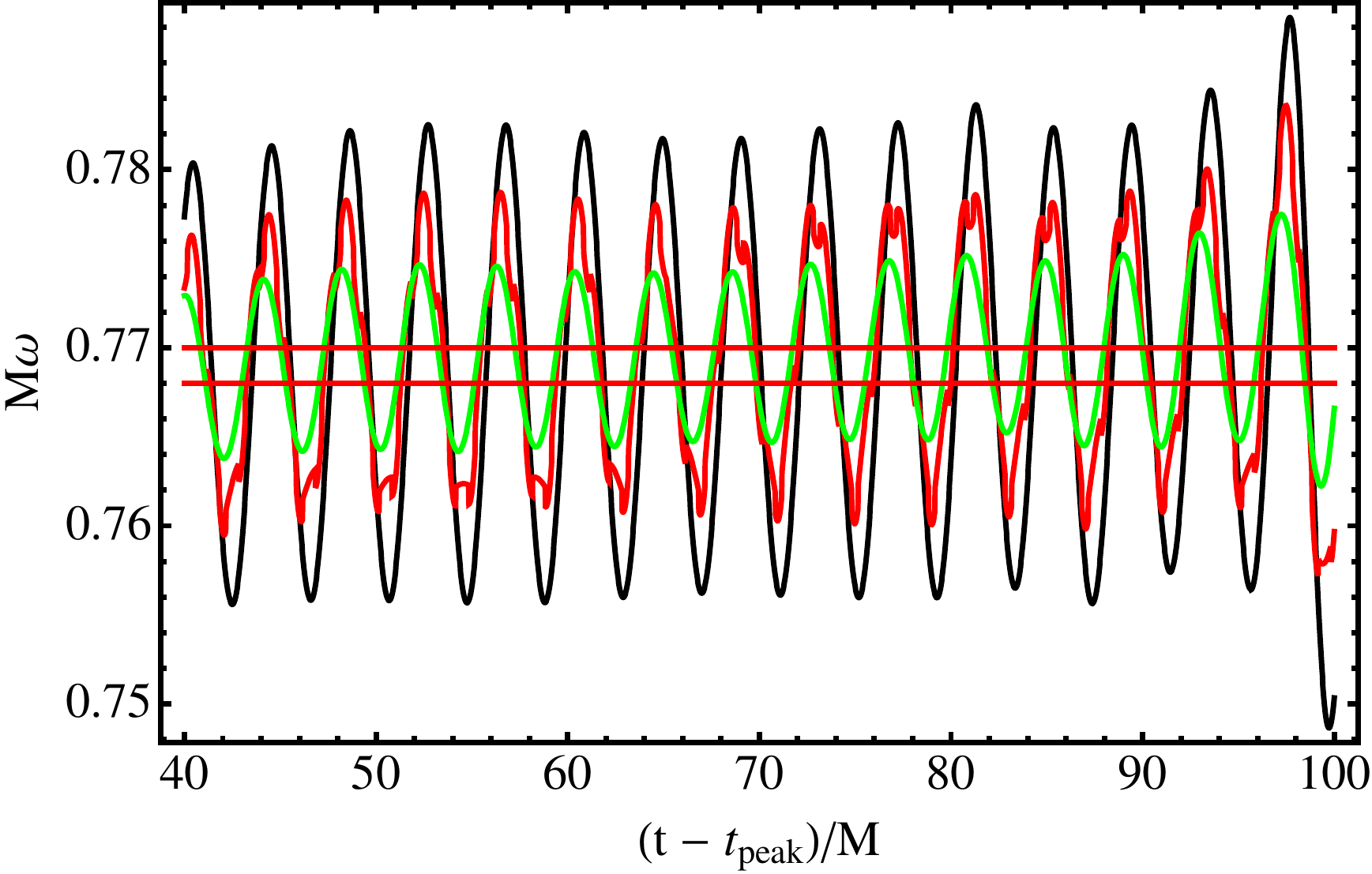}
\caption{
The numerical GW frequency at a time $(t-t_{\rm peak})/M$ after the peak of $|r\Psi_{4,22}|$, shown for
simulations at three resolutions. The frequency oscillates around a value that we take to be the 
ringdown frequency. The amplitude of the oscillations decreases as the numerical resolution is 
improved, suggesting that these are only a numerical artifact. The upper and lower bounds of 
our estimate of the ringdown frequency are indicated by the two horizontal lines in the plot. The 
bounds were obtained by considering the results from all three numerical resolutions, and varying
the portion of the data used for the fit; the final quoted values were calculated using the range 
$(t-t_{\rm peak})/M \in \{50,100\}$.  Despite the high amplitude of the noise in the data,
the average value shows very little variation, and can be estimated with an uncertainty of only 
$\Delta \omega = 0.001/M$. 
}
\label{fig:RingdownFreq}
\end{figure}

In the $\chi_i = 0.85$ case, we find that the ringdown frequency is $M\omega_{\rm RD} = 0.769 \pm 0.001$;
see Fig.~\ref{fig:RingdownFreq}.
The final mass is $M_f/M = 0.895 \pm 0.015$, and the final spin is $a_f/M_f = 0.915 \pm 0.007$. 
Note that the final mass is lower than quoted in~\cite{Hannam:2007wf}, where all of the analysis
was performed on the highest-resolution waveform calculated on the largest radiation extraction sphere. 
Here we extrapolate the results with respect to extraction radius (assuming a $1/R_{\rm ex}$ fall-off
in the error), and with respect to numerical resolution, where the results show between 2nd- and
6th-order convergence. The radiated energy increases with extraction radius, and so our estimate of the final
mass decreases; this is why our extrapolated value (0.895) is lower than the $R_{\rm ex} = 90M$ 
value of 0.911 quoted in~\cite{Hannam:2007wf}. In addition, we estimate the ringdown frequency 
using a $50M$-long sample of the waveform starting $50M$ after the peak amplitude of the 
$(\ell=2,m=2)$ mode, while our earlier results were based on a portion of the waveform starting only a few 
$M$ after the peak amplitude, which distorts the final estimate of the ringdown frequency. 

For comparison, Dain, {\it et. al.}~\cite{Dain:2008ck} study the $\chi_i = 0.92$ case. The initial black-hole
spins are larger than studied here (and were set up to approach the highest spin possible for 
Bowen-York data), and therefore the final black holes should have a larger ringdown frequency than in our
$\chi_i = 0.85$ case, and a higher final spin, and this is indeed the case. Note that they use the ADM mass 
$M_{\rm ADM}$ as the defining length scale in their simulations, while we use the total BH mass. In
terms of the ADM mass, the ringdown frequency for the $\chi_i = 0.85$ case is 
$M_{\rm ADM} \omega_{\rm RD} = 0.761$, while the frequency found in~\cite{Dain:2008ck} for the $\chi_i = 0.92$
case is $M_{\rm ADM} \omega_{\rm RD} = 0.766$.
In addition, they give a measure of the final spin between 0.910 and 0.916, 
where 0.915 is the value obtained using the same quasi-normal-mode method that we have applied here. 
Their final-spin result is consistent with ours'  within our error bounds.

A number of aligned-spin cases were studied in~\cite{Marronetti:2007wz}. For the cases where direct
comparison is available, our results show agreement within 1\% for spins up to 0.5, and within 2\% for
higher spins. Assuming that their uncertainties are comparable to ours', then the results agree. The
first hang-up and anti-hang-up cases were studied in~\cite{Campanelli:2006uy}, each with spin values
of $|\chi_i| = 0.757$. They estimate final spins of 0.443 and 0.890 for the anti-hangup and hangup cases
respectively, and these are also consistent with our results. Two of the unequal-mass cases 
were also studied in~\cite{Baker:2008mj}, $q=2,4$, and the final mass and spin results are in excellent
agreement. 

We have also compared our results with fits for the  final spin available in the literature. 
We find excellent agreement to about 1 \% or better with
\cite{Rezzolla:2007rd}, as well as with \cite{Lousto:2009mf} 
as long as the spins are not anti-aligned with the orbital angular momentum.
For the  latter paper we find disagreements of $\approx 10 \%$ for
the cases  $\chi_i = -0.75, -0.85$.

\begin{table*}
\caption{\label{tab:configurations}
Summary of the configurations simulated. The table indicates the initial coordinate separation $D/M$
of the punctures, their tangential and radial momenta ($p_t,p_r$), and the eccentricity $e$ of the resulting
coordinate motion. For the $\chi_i < 0$ cases, where enhanced PN parameters were used to achieve 
low-eccentricity inspiral, the eccentricity from raw PN-inspiral parameters is also shown in brackets.
The initial GW frequency is $M\omega_i$, and the ringdown frequency of the final
merged black hole is $M\omega_{\rm RD}$. The simulation includes $N_{\rm GW}$ cycles before
the peak of the GW amplitude, which occurs at $t_{\rm peak}$. The final black hole has mass $M_f$
and spin $a_f$, and receives a recoil of $v_{\rm kick}$.} 
\begin{tabular}{|| c | c | c | c | c | c | c | c | c | c | c | c | c ||}
\hline
$q$  &  $S_i/M_i^2$  &  $D/M$ & $p_t/M$    & $-p_r/M (\times 10^{-4})$    & $e$     & $M\omega_i$ & $N_{\rm GW}$ &  $t_{\rm peak}/M$ & $M\omega_{\rm RD}$ & $M_f/M$ &  $a_f/M_f$ & $v_{\rm kick}$ (km/s) \\
\hline
1       &   -0.85              & 13.0      & 0.084542 & 5.247                                    & 0.0025 (0.009) &  0.040   & 16    & 1868 &  0.457 & 0.969   &  0.412   & 0  \\
1       &   -0.75              & 13.0      & 0.084057 & 5.060                                    & 0.0016 (0.008) &  0.0395 & 17    & 2036 & 0.466 & 0.968   &  0.446   & 0 \\
1       &   -0.50              & 12.5      & 0.085124 & 5.258                                    & 0.0029 (0.0045) &  0.042   & 18    &  2065 & 0.490 & 0.965   &  0.531   & 0  \\
1       &   -0.25              & 12.0      & 0.086312 & 5.623                                     & 0.0025 (0.004) &  0.044   & 18.5 &  1955 & 0.519 & 0.959   &  0.609   & 0  \\
1       &    0                    & 12.0      & 0.085035 & 5.373                                    & 0.0018 &  0.044   & 19    &  1939 & 0.553 & 0.951   &  0.686    &  0  \\
1       &   0.25               & 12.0      & 0.083813 & 0                                            & 0.0061 &  0.043    & 21.5 &  2129 & 0.595 & 0.942   &  0.760   & 0 \\
1       &   0.50               & 11.0      & 0.087415 & 0                                             & 0.0061 &  0.049    & 20    &  1739 & 0.650 & 0.936   &  0.832   & 0  \\
1       &   0.75               & 10.0      & 0.091435 & 0                                            & 0.0060 &  0.055    & 19    &  1432 & 0.728 & 0.921   &  0.898   & 0 \\
1       &   0.85               & 10.0      & 0.090857 & 0                                             & 0.0050 &  0.055    & 20    &  1492 & 0.770 & 0.895   &  0.915   & 0  \\
\hline
2       &  0            & 10.0               & 0.085599  & 7.948                                   & 0.0023 & 0.058      & 12.5  &  1069 & 0.522 & 0.962   &   0.623  & $140 \pm 5$ \\
3       &  0            & 10.0               & 0.072408  & 5.802                                    & 0.0016 & 0.058      & 14.5  &  1240 & 0.489 & 0.972   &   0.540  & $155 \pm 15$  \\
4       &  0            & 10.0               & 0.061914  & 4.333                                  & 0.0038 & 0.056      & 17     &  1461 & 0.467 & 0.978   &   0.471  & $145 \pm 10$  \\
4       &  0            & 10.0               & 0.061883  & 4.211                                  & 0.0026 & 0.057      &  16    & 1396  & 0.467  & 0.978  &  0.471  & $145 \pm 10$ \\
\hline
\end{tabular}
\end{table*}

\subsection{Energy spectrum in spherical harmonic modes}
\label{sec:highermodes}

The $(\ell=2,m=2)$ mode dominates the GW signal from a black-hole-binary coalescence,
and indeed most current searches in detector data employ templates that include only this 
harmonic \cite{Abbott:2009qj,Abbott:2009tt} (see also the NINJA project
searches in simulated data with injected numerical relativity waveforms 
\cite{Aylott:2009tn,Aylott:2009ya}). However, knowledge of the subdominant modes may aid detection, and 
are important for accurate estimation of the source 
parameters~\cite{Sintes:1999cg,VanDenBroeck:2006ar,Babak:2008bu,Thorpe:2008wh,McWilliams:2009bg}.

We assess the relative importance of the sub-dominant modes by calculating the 
energy radiated in each mode. The radiated energy in each mode is given by~\cite{Alcubierre2008} 
\begin{equation}
\frac{dE_{\ell,m}}{dt} = \lim_{r \rightarrow \infty} \frac{r^2}{16\pi} \left| \int_{-\infty}^t \Psi_{4,\ell m} dt' \right|^2.
\end{equation}
In practice the limits of the integration are taken as the time in the simulation just after the 
junk radiation has passed, and a time after the signal has rung down to the level of numerical
noise.
The results are summarized in Tab.~\ref{tab:modeenergy}, including only those
modes that contribute above 1\% of the total energy.
We see that in the equal-mass cases, the 
$(2,2)$ mode  dominates --- around 98\% of the energy is radiated in the dominant
mode in all cases, with only a negligible variation due to spin, and no other modes contribute above 1\%.

In the unequal-mass cases, the energy contribution from the higher harmonics grows rapidly with mass
ratio, particularly in the $\ell = \pm m$ modes. We defer the reader to the detailed discussion 
in~\cite{Berti:2007fi}, but note that, even at $q=4$, most of the energy is radiated in a very small number 
of harmonics.

\begin{table*}
\caption{\label{tab:modeenergy}
Ratio of total energy radiated in each mode. Only contributions above 1\% are included. 
}
\begin{tabular}{|| c | c | c | c | c | c ||}
\hline
Case                    & $(2,\pm2)$                &  $(2,\pm1)$              & $(3,\pm3)$ & $(4,\pm4)$               & $(5,\pm5)$       \\
\hline
$\chi_i = -0.85$ & 0.988       & 0 &  0 &  --- & 0  \\
$\chi_i = -0.50$ & 0.989       & 0 &  0 &  --- & 0  \\
$\chi_i = 0       $ & 0.990       & 0 &  0 &  --- & 0  \\
$\chi_i = 0.50 $ & 0.988       & 0 &  0 &  --- & 0  \\
$\chi_i = 0.85 $ & 0.988       & 0 &  0 &  --- & 0  \\
\hline
$q=2$                 & 0.947       & --- & 0.038 &  --- & ---  \\
$q=3$                 & 0.897       & --- & 0.076 & 0.013 & --- \\
$q=4$                 & 0.868       & 0.013 & 0.095 & 0.017 & --- \\ 
\hline
\end{tabular}
\end{table*}

\subsection{Recoil}
\label{sec:recoil}

Due to the asymmetry of the radiation emission in the unequal-mass cases, 
linear momentum is radiated from the system, and the center-of-mass of the 
binary moves as the black holes inspiral. The direction of the center-of-mass recoil 
rotates with the binary, so that the average movement is small.
However, the rate of momentum loss grows as the black holes get closer, and, 
as with the total GW signal, peaks at merger. This final burst of GW emission 
causes an overall recoil, or ``kick''.

Since the bulk of the recoil arises during the merger, short simulations are sufficient
to accurately measure the effect, and these were used 
in~\cite{Baker:2006vn,Gonzalez:2006md,Herrmann:2007zz} to make the first 
accurate fully general-relativistic predictions of gravitational recoil, and found that 
the maximum kick for nonspinning binaries is $v_{\rm max} = 175\pm11$\,km/s for a mass ratio 
of $q = 2.8$~\cite{Gonzalez:2006md}.
An analytical fitting formula for the recoil from nonspinning binaries was 
presented in~\cite{Gonzalez:2006md}, for recent papers containing such
fitting formulas see \cite{Baker:2008md} (which uses the
same ansatz as \cite{Gonzalez:2006md} and finds slightly different but
consistent fitting parameters) 
and \cite{Lousto:2009mf} (which quotes
the fit from~\cite{Gonzalez:2006md} for nonspinning binaries).

Tab.~\ref{tab:configurations} shows the results for the current simulations, which 
agree with those from the shorter simulations presented in~\cite{Gonzalez:2006md}.
It has also been shown that much larger recoils are possible from spinning 
or highly elliptical
binaries~\cite{Koppitz:2007ev,Choi:2007eu,Campanelli:2007ew,Gonzalez:2007hi,Campanelli:2007cga,Brugmann:2007zj,Pollney:2007ss,Lousto:2007db,Dain:2008ck,Healy:2008js},
but not for any of the configurations that we have studied in this work.

Our newer simulations improve over those produced in~\cite{Gonzalez:2006md} in
two ways: they include many more cycles before merger, and the wave extraction
is performed at larger radii. On the other hand, the numerical resolution at the 
wave extraction radii is lower, which reduces the accuracy. As such, the values we
quote in Tab.~\ref{tab:configurations} have large error bars. 

The flux of angular momentum radiation is given by 
\beq
\frac{dP_i}{dt} = \lim_{r \rightarrow \infty} \left[ \frac{r^2}{16\pi}
  \int_{\Omega} \ell_i \left| \int_{-\infty}^{t} \Psi_4 d\tilde{t}
  \right|^2 d{\Omega}\right]. 
  \label{eqn:Pint}
\eeq
where $ \ell_i = \left(\sin \theta \cos \phi, \sin \theta \sin \phi, 
\cos \theta \right)$~\cite{Campanelli:1998jv}. The total recoil is
calculated by integrating Eqn.~(\ref{eqn:Pint}) over the duration of the 
simulation. 

The additional length of the new simulations allows us to remove one source of error in
shorter simulations: the choice of starting time in the 
 integration of $dP_i/dt$ to calculate the total radiated linear momentum. This function 
 oscillates with time, and during the inspiral the average radiated linear momentum is 
 much smaller than the amplitude of the oscillations --- so a poor choice of starting time
 in the integration of $dP_i/dt$ could potentially corrupt the final result. 
 In~\cite{Gonzalez:2006md} the uncertainty due to this effect was estimated at about 3\%. 
 In~\cite{Baker:2006vn,Koppitz:2007ev} attempts were made to both account for this
 effect and for the linear momentum loss that will have accumulated over the earlier
 inspiral of the binary. In our cases, where we possess the waveform for many more
 cycles before merger, we are able to simply calculate the total recoil for a range of 
 integration starting times $t_0$, and then to take the average of these values. We find that
 the uncertainty in this process is only a fraction of a percent of the final result. 
 
\begin{figure}[t]
\centering
\includegraphics[width=80mm]{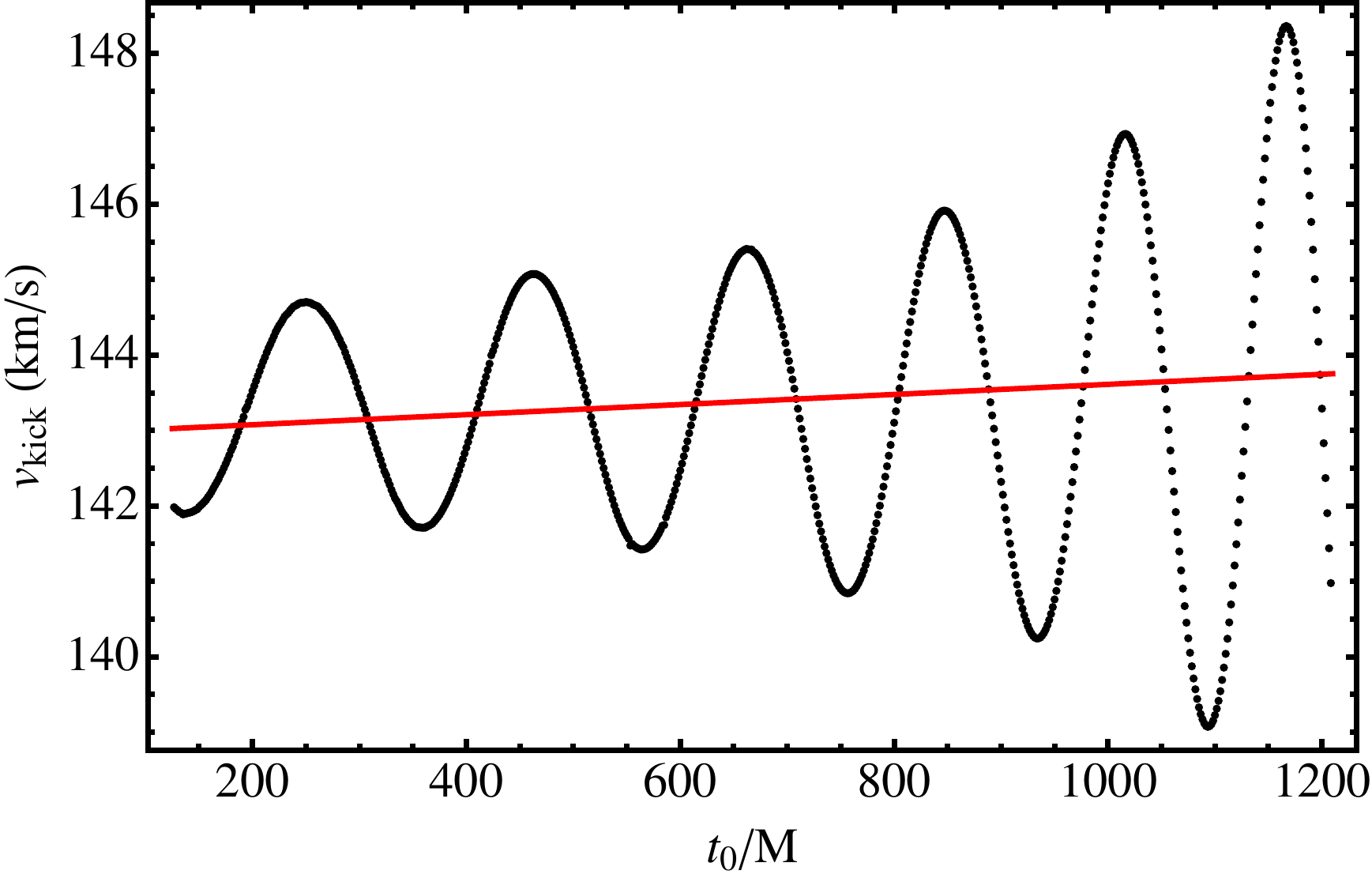}
\caption{
Variation in the estimate of the total radiated linear momentum (recoil),
as a function of the starting time $t_0$ of the integration of $dP_i/dt$ for
the $q=4$ case. 
The value oscillates around a slowly growing average, which is indicated
by a straight line. Note that the latest integration time used, $t_0 = 1200M$, 
is approximately $300M$ before the end of the ringdown, which is the
point at which the integration was started in the older calculations 
in~\cite{Gonzalez:2006md}.
}
\label{fig:KickVariation}
\end{figure}

 Figure~\ref{fig:KickVariation} illustrates this effect with the $q=4$ case. The lower limit
 of the integration, $t_0$, was varied between $t_0 = 125M$ (just after the burst of 
 junk radiation has passed through the signal, and $t_0 = 1200M$, which is about
 $300M$ before $dP_i/dt$ has fallen to negligible values, and also roughly 
 corresponds to the value of $t_0$ that was used for the much shorter waveforms
 studied in~\cite{Gonzalez:2006md}. Note that the kick calculated for different 
 choices of $t_0$ varies by about 4\,km/s, or 3\% of the result. A linear curve fit
 through the results (shown in the figure) indicates that the average result of the 
 integrated linear momentum radiation rises very slowly during the inspiral,
 and varies by only 0.7\,km/s, or 0.5\%. In our results, we determine the final
 kick to be the average over this range of choices $t_0$, which introduces only a 
 negligible error in our result.
 
 A second error source that could not be quantified
 in~\cite{Gonzalez:2006md} was that due to extraction of the GW signal at finite 
 extraction radius. In that work an extraction radius of only $R_{\rm ex} = 30M$ was
 feasible. We now extract GW signals at up to $R_{\rm ex} = 90M$, although we find
 that the numerical resolution at the wave extraction spheres allows an accurate
 calculation of the recoil only for $R_{\rm ex} = \{50,60\}M$. However, these two 
 radii are sufficient for us to extrapolate the recoil to $R_{\rm ex} \rightarrow \infty$,
 assuming a $1/R_{\rm ex}$ fall-off in the error. This gives the values listed in 
 Tab.~\ref{tab:configurations}. This fit also of course predicts the value of the 
 recoil at $R_{\rm ex} = 30M$, which agrees well with the values in~\cite{Gonzalez:2006md}.
 However, due to the poorer numerical resolution on the extraction spheres, we
 assign large error bars to our values.
 
\section{Late inspiral comparison of NR and PN waveforms}
\label{sec:pncomparison}

One of the most important applications of our waveforms is as input in the construction 
of analytic waveform models that can in turn be used to construct template banks for
GW searches. In particular, these waveforms have already been used to produce the 
phenomenological models presented 
in~\cite{Ajith:2007kx,Ajith:2007xh,Ajith:2009bn,Santamaria:2010yb}, in which the 
NR waveforms for the late inspiral and merger are connected to long
PN inspiral waveforms, to produce ``complete'' waveforms for the full inspiral-merger-ringdown,
and it is these complete waveforms that are then used in the construction of a 
phenomenological model that is essentially an analytic fit across the relevant section of the 
black-hole-binary parameter space. 

In performing this procedure, we need to quantify the level of agreement between the 
PN and NR waveforms in some region where they are both considered to be valid. 
In other words: the approximate PN waveforms are expected to be accurate during much of 
the long inspiral, but are they still accurate enough at the point where we want to connect them to 
fully general relativistic results? 

This question was first addressed for equal-mass nonspinning waveforms 
in~\cite{Buonanno:2006ui,Baker:2006ha},
and later with increasing levels of precision in~\cite{Hannam:2007ik,Boyle:2007ft,Gopakumar:2007vh}.
The conclusion of these works was that the phase disagreement between NR waveforms and 
typical PN approximants was less than 1\,rad over the last 10 cycles up to $M\omega = 0.1$,
and the error in the quadrupole PN amplitude was about 8\%; however, the phase could be
tracked with surprising accuracy by one PN approximant, TaylorT4, and the PN amplitude was
accurate to within 2\% when evaluated at 3PN order~\cite{Boyle:2007ft,Kidder:2007rt}. 

Similar
comparisons were performed with the equal-mass $\chi_i > 0$ cases that we consider here,
where it was found that the phase disagreements were comparable for all spin values for 
the TaylorT1 approximant (i.e., approximately 1\,rad over the 10 cycles up to $M\omega = 0.1$), 
but that the TaylorT4 approximant, which performed so well in the nonspinning case, did no
better than TaylorT1 (and often worse) when the BHs were spinning. It should be noted,
however, that the Taylor approximants did not include spin terms up to the same
PN order as in nonspinning terms (2.5PN versus 3.5PN), which is a point we will return to later. 
In addition, it was found that  the quadrupole amplitude error grew to as much as 12\% in high 
spin cases~\cite{Hannam:2007wf}. 
Equal-mass nonspinning eccentric binaries were considered in~\cite{Hinder:2008kv}, and
one unequal-mass precessing-spin configuration was studied in~\cite{Campanelli:2008nk}.

In this section we will perform a PN-NR comparison for all of our waveforms, 
which now include $q\neq 1$
and $\chi_i < 0$ cases.

\subsection{PN approximants}

The PN approximants considered here are derived from the energy $\mathcal E$ and 
GW flux $\mathcal F$ of a black-hole binary on quasi-circular orbits. Both quantities 
are given in the PN framework as expansions in $v/c$, up to $(v/c)^7$ (3.5PN order), 
where $v$ is the relative velocity and $c$ the speed 
of light. Following the standard convention, we regard $\mathcal E$ and $\mathcal F$ as 
functions of the dimensionless
variable $x = (v/c)^2$ that is related to the orbital phase $\phi_{\rm orb}$ via
\begin{align}
 x &= (M\omega_{\rm orb})^{2/3}~, & \frac{d\phi_{\rm orb}}{d t} = \omega_{\rm orb} ~. \label{eq:xomega}
\end{align}
The energy-balance law $d\mathcal{E}/dt = - \mathcal F$
can be transformed to an evolution equation for $x$,
\begin{equation}
\frac{dx}{dt} = - \frac{\mathcal F}{d \mathcal E/dx} \label{eq:dxdt}
\end{equation}
which in turn leads to the $\ell m$-mode of the gravitational wave strain
\begin{equation}
 h^{\ell m} (t) = H^{\ell m} (t) \, e^{- i m \phi_{\rm orb}(t)}~.
\end{equation}
The amplitudes $H^{\ell m}$ are given as expansions in $x$ to 3PN order in the non-spinning
case \cite{Blanchet:2008je} and up to 2PN order in spinning contributions \cite{Arun:2008kb}.

A direct (numerical) integration of (\ref{eq:dxdt}) and (\ref{eq:xomega}) is referred to as the 
\emph{TaylorT1} approximant. If instead the right-hand side of Eq.~(\ref{eq:dxdt}) is re-expanded 
as a Taylor series in $x$ before integrating, the resulting approximant is called \emph{TaylorT4}. 
This re-expansion is truncated at the same order as the energy and flux (i.e., 3.5PN); all higher powers 
in $x$ are incomplete and therefore neglected.

If we apply the same strategy to the spin contributions that enter at 1.5PN (leading order 
spin-orbit coupling),  2PN (spin-spin) and 2.5PN order (next to leading order spin-orbit), 
we should neglect all spin-dependent terms in the re-expansion of (\ref{eq:dxdt}) that appear 
at 3PN and 3.5PN order. We denote the resulting approximant that was used for instance 
in~\cite{Hannam:2007wf} as \emph{TaylorT4 (truncated)}. If we instead disregard the distinction 
of spinning and non-spinning terms and use the ``full'' re-expansion up to 3.5 PN order, 
thereby keeping incomplete spin contributions at 3 and 3.5PN order, we denote the 
resulting approximant simply as \emph{TaylorT4}. For a detailed discussion and explicit 
expressions for the approximants see \cite{Santamaria:2010yb} and references therein. 
Further small corrections to the spin contributions to the PN phase and amplitude, due to typographical or other errors in the original literature, were found during a program of 
PN-approximant verification within the Ninja collaboration~\cite{ninja-wiki}; these will
be described in more detail in an upcoming amendment to the data format specification 
document~\cite{Brown:2007jx}, and are discussed further in Sec.~\ref{sec:ampcomparison}
below.

\subsection{Phase comparison} 

We now compare the PN and NR phase. Our procedure, as in previous 
studies~\cite{Hannam:2007ik,Hannam:2007wf}, is to consider the phase for the 
$N$ GW cycles up to the matching frequency $M\omega_m = 0.1$. We line up the PN and 
NR phase functions
so that they agree when $\omega = \omega_m$, and relabel this event as $t=0$. 
We then calculate the phase disagreement as it accumulates over $N$ cycles
back in time. Note that although our comparison is over a fixed number of
GW cycles, it is \emph{not} over a fixed frequency range, due to the different frequency 
evolution in each configuration. In the same way, the comparison is also over different
lengths of time between different configurations. However, we have found that 
the qualitative behaviour of the comparison results does not depend on whether
we compare over a fixed range of cycles, frequency, or time.

In previous studies we simply calculated the phase difference
$\Delta \phi(t) = \phi_{\rm PN}(t) - \phi_{\rm NR}(t)$, and quoted $\Delta \phi(t_{N})$
as the accumulated phase difference, where $t_{N}$ is the time $N$ cycles 
prior to the point where $\omega = \omega_m$. This procedure gives consistent
results, but we may worry in general that $\Delta \phi(t)$ is not a monotonic 
function, and so a more robust procedure is to consider instead \begin{equation}
\overline{\Delta \phi}(t_{N}) =  \frac{1}{\sqrt{-t_{N}}} \left[ \int_{t_{N}}^{0} \Big( \phi_{\rm NR}(t) - \phi_{\rm PN}(t) \Big)^2 dt \right]^{1/2}. \label{eqn:sqrdphi}
\end{equation} This gives us a measure of the average rate of increase of the 
phase disagreement. A similar procedure was also used in~\cite{Boyle:2008ge}, 
although in that study the alignment of the waveforms was adjusted to 
minimize $\overline{\Delta \phi}$. An elegant alternative measure of the accumulated phase
disagreement is given in Eqn.~(3.15) of~\cite{McWilliams:2008}.
We instead wish to evaluate how well the PN phase evolution
agrees with the fully general relativistic NR results.
For comparison with previous results in the literature, we will also show the results of
a direct calculation of $\phi_{\rm PN}(t) - \phi_{\rm NR}(t)$.

\begin{figure}[t]
\centering
\includegraphics[width=83mm]{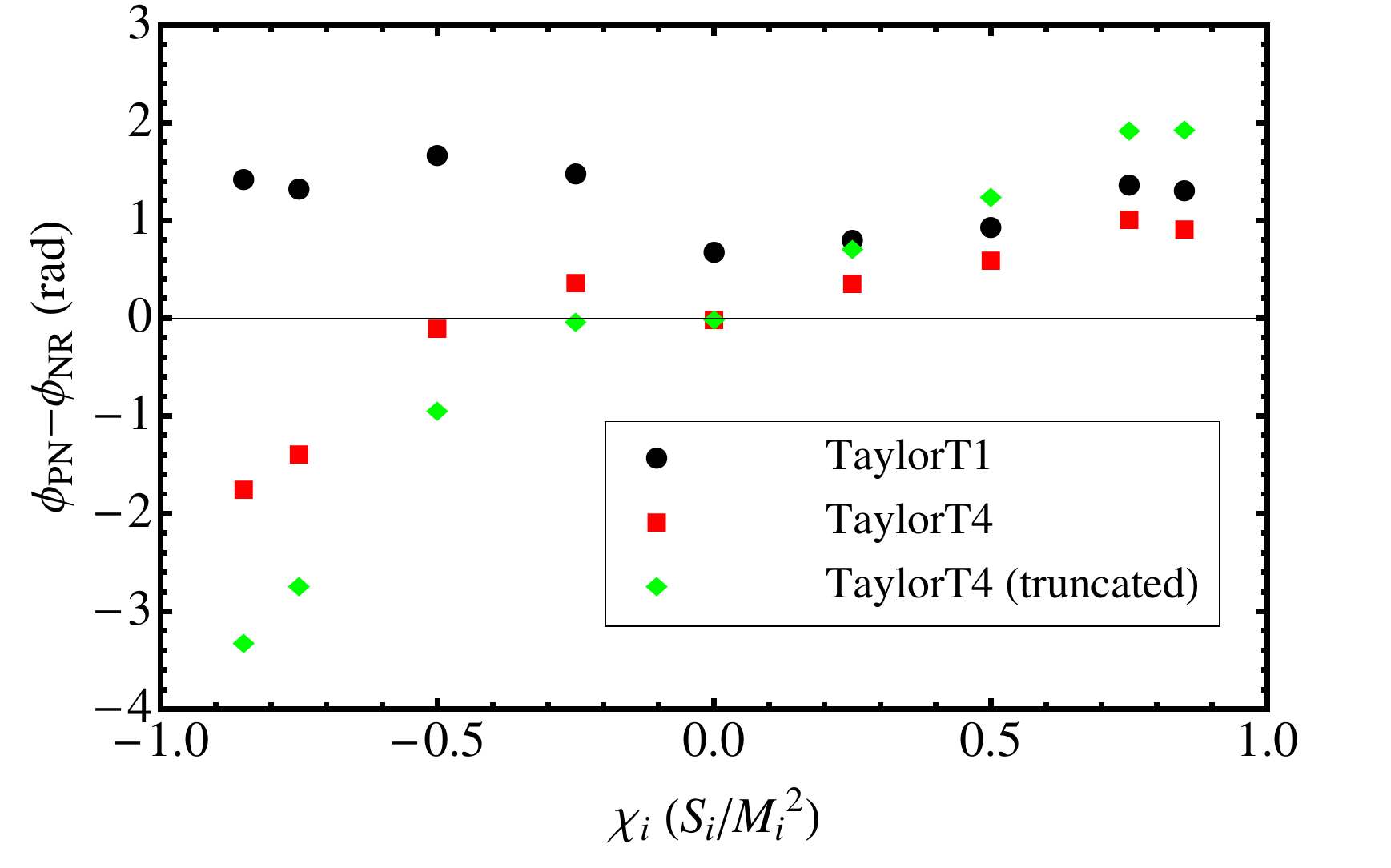}
\includegraphics[width=80mm]{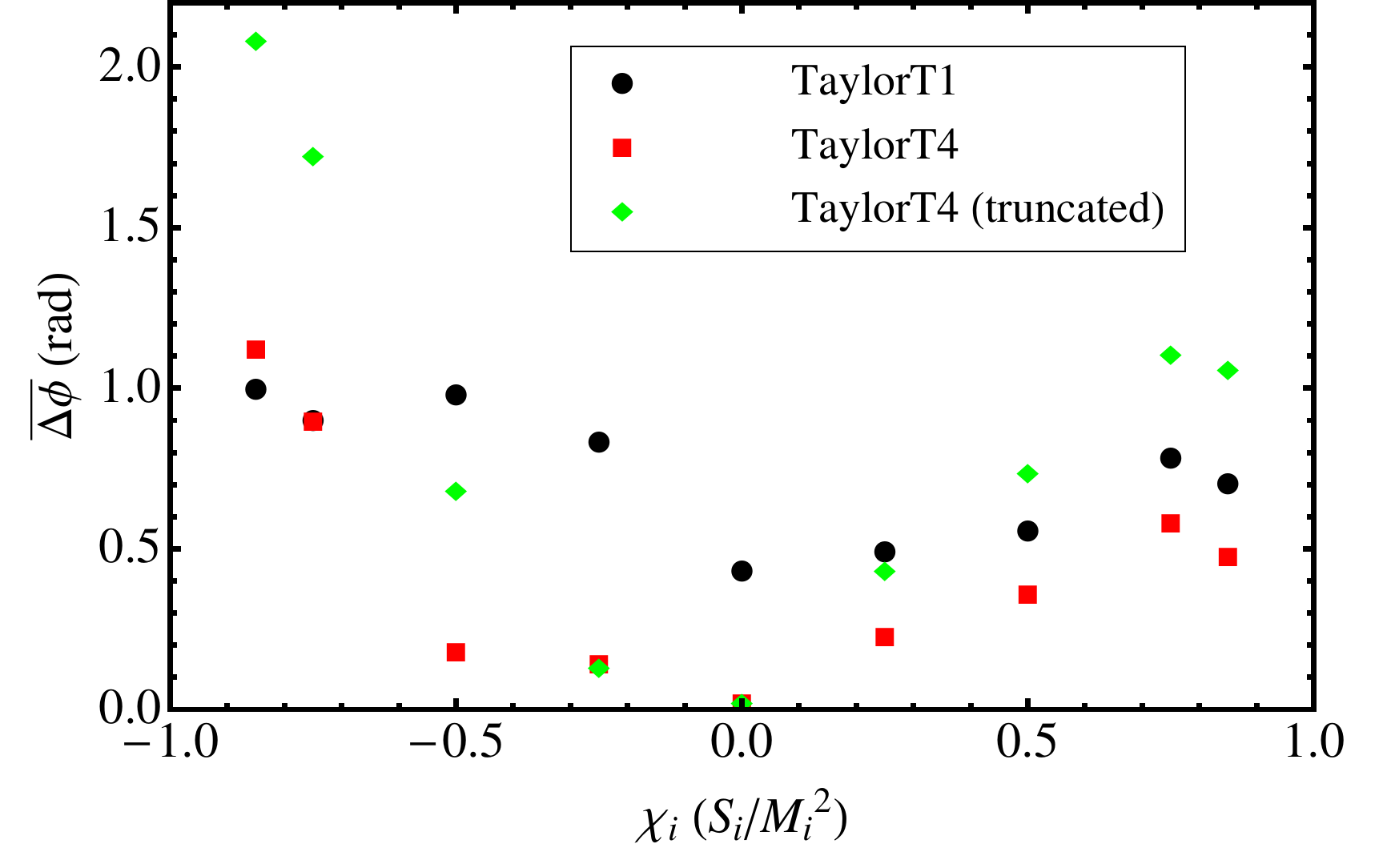}
\caption{
Phase disagreement between NR and PN results for three choices of PN
approximant, for configurations that consist of equal-mass binaries with equal
spins oriented parallel or anti-parallel to the orbital angular momentum.
The first panel shows the accumulated phase disagreement for the ten GW 
cycles up to $M\omega_m = 0.1$. The second panel shows the integrated 
square of the phase disagreement, Eqn.~(\ref{eqn:sqrdphi}).
}
\label{fig:PNSpinPhaseComp}
\end{figure}

Fig.~\ref{fig:PNSpinPhaseComp} shows the disagreement between the PN and
NR phase for the equal-mass configurations with non-precessing spins over $N = 10$
GW cycles. Three
PN approximants are used: TaylorT1, TaylorT4, and TaylorT4-truncated, as described in
the previous section. 

We see that in both calculations of the accumulated phase disagreement, TaylorT1 
is the most robust. It performs best in the nonspinning case (which is to be expected, since
the nonspinning contributions are known to higher PN order than the spinning contributions),
and for all spinning cases the accumulated phase disagreement is between 1.0 and 2.0\,rad, while
the square-averaged phase disagreement is between 0.5 and 1.0\,rad. We see also that
TaylorT4-truncated performs worse as the spin is increased, and for large anti-aligned spins
performs very poorly. The full TaylorT4 approximant performs better for most spin values, although it
is again poor for large anti-aligned spins. It is in light of comparisons using only TaylorT1 and
TaylorT4-truncated that we chose to use the TaylorT1 approximant in the construction of 
hybrid waveforms for the phenomenological model in~\cite{Ajith:2009bn}.

\begin{figure}[t]
\centering
\includegraphics[width=82mm]{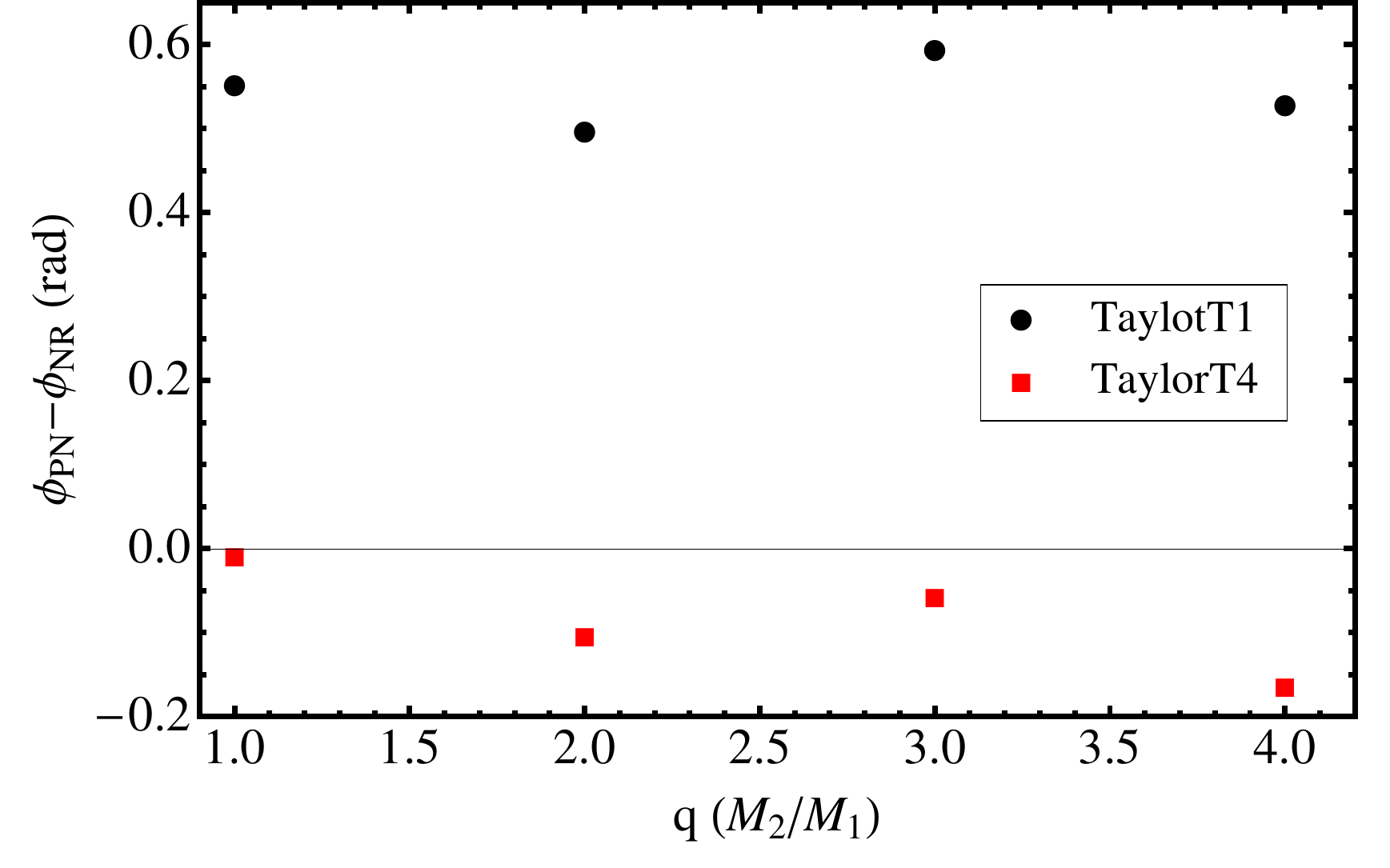}
\includegraphics[width=80mm]{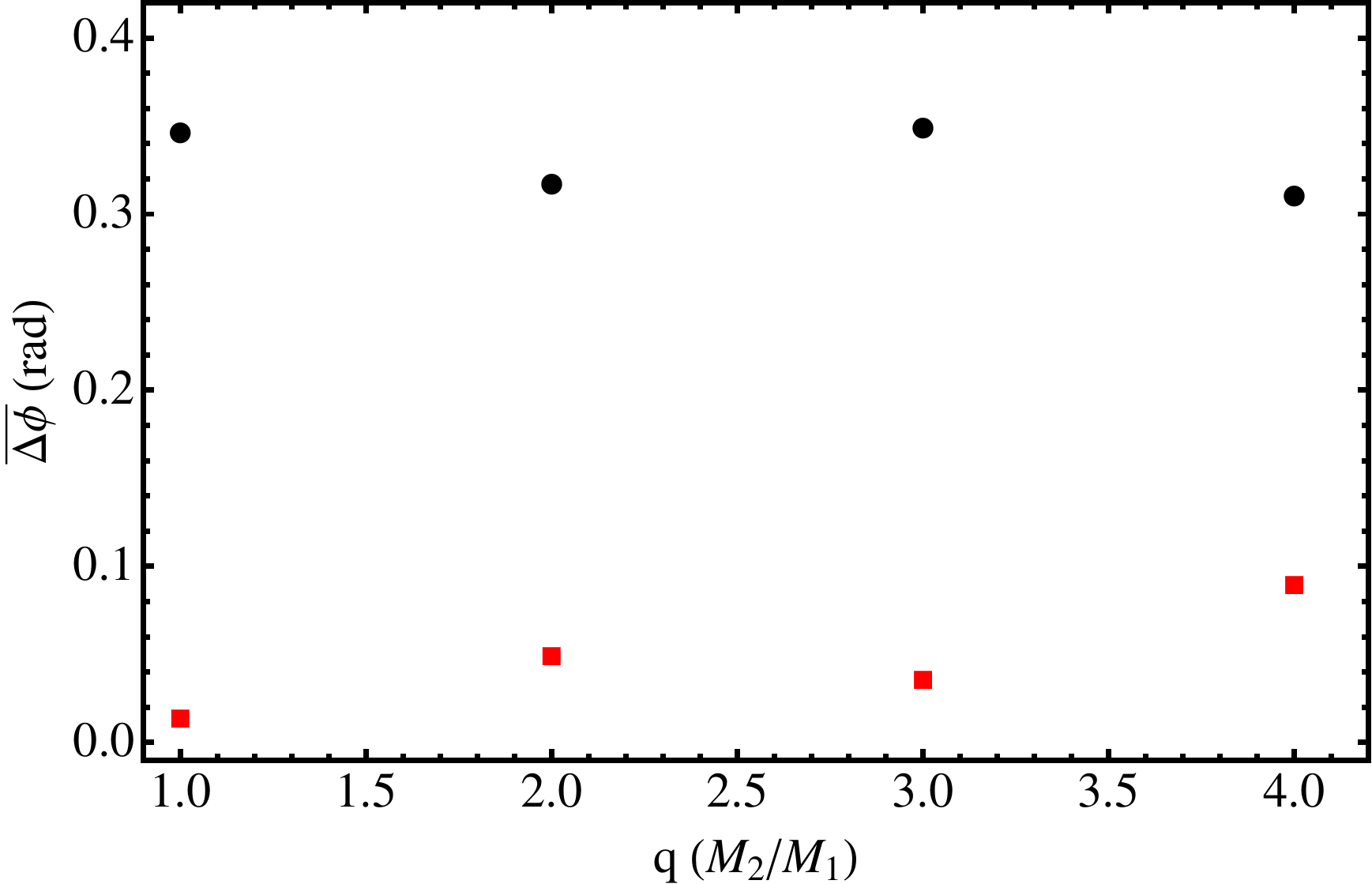}
\caption{
Phase disagreement between NR and PN results for two choices of PN
approximant, for configurations that consist of nonspinning black holes of 
unequal mass, with mass ratio $q = M_2/M_1$.
}
\label{fig:PNUMPhaseComp}
\end{figure}

Fig.~\ref{fig:PNUMPhaseComp} shows a similar plot, but this time for the unequal-mass
nonspinning configurations. The $q=2$ simulations consist of less than ten cycles before 
$M\omega = 0.1$, so we consider only $N=8$ cycles in the phase comparison. 
In this case we see that TaylorT4 continues to perform well
for unequal-mass configurations. We expect that at higher mass ratios the performance
of all PN approximants will deteriorate, but up to $q=4$ this deterioration 
cannot be clearly measured; the performance of TaylorT1 and TaylorT4 shows some
variation with mass ratio, but this is not monotonic.

From our phase comparison analysis, we conclude that the TaylorT1 approximant is
most robust over the entire subset of the black-hole-binary parameter space that we have
studied. The TaylorT4 approximant performs well for all nonspinning cases. The 
performance of TaylorT4 for spinning cases varies greatly between our two choices of 
treatment of the higher-order spin contributions, but for both choices shows poor agreement
for large anti-aligned spins. We caution, however, that the performance of the approximants
over a relatively small number of numerical cycles does not tell us how well they perform
before at lower frequencies, and we will return to this point in the Discussion.

\subsection{Amplitude comparison}
\label{sec:ampcomparison}

We now compare the PN prediction for the inspiral wave amplitude with numerical results,
for the $(\ell=2,m=2)$ mode. We found in~\cite{Hannam:2007ik} that in the equal-mass
nonspinning case the quadrupole PN amplitude was larger than the full GR amplitude during
inspiral by about 7\%. It was later shown in~\cite{Boyle:2007ft} that the amplitude agreement
could be improved to within 2\% if corrections up to 3PN order were used. For equal-mass
binaries with aligned spins, we found in~\cite{Hannam:2007wf} that the quadrupole PN 
amplitude disagreement rose to about 12\% in highly spinning cases. 

In this section we extend our previous analysis of the quadrupole amplitude 
to anti-aligned and unequal-mass cases. We also compare with the PN amplitude that
results from using all currently known amplitude corrections (up to 3PN order 
non-spinning~\cite{Kidder:2007rt,Blanchet:2008je} and up to 2PN order spinning 
contributions~\cite{Will:1996zj,Arun:2008kb}). We have taken care when combining results 
for amplitude functions from different sources in the  literature,
in particular regarding different conventions for the choice of relative phase factors. 
In our implementation we now follow the convention of~\cite{Arun:2008kb}, 
which differs from that of~\cite{Blanchet:2008je}, from which we originally took our 
nonspinning amplitude contributions. 
We have checked for consistency with the amplitude of the $l=\vert m \vert =2$ 
modes as given in~\cite{Berti:2007nw}, and we have compared with an independent 
code as part of the Ninja project ~\cite{ninja-wiki,boyleprivate}. In addition,
we have also checked that inconsistent choices of the relative 
phase factors  (e.g., caused by misprints in the literature) significantly increase 
the deviation of the NR and PN amplitudes; the correct choices lead to the 
best agreement with results from full general relativity. 

We find that the GW amplitude shows variations with numerical extraction radius that are comparable to the 
level of disagreement with the PN predictions. However, the error in the amplitude seems to fall
off as $1/R_{\rm ex}^2$ (see~\cite{Hannam:2007ik} for a discussion of this effect), and allows us to 
perform an accurate extrapolation to $R_{\rm ex} \rightarrow \infty$. Having obtained the accurate
amplitude of $R_{\rm ex} \Psi_4$, we then express the amplitude as a function of frequency, using
the methods we introduced in Sec.~\ref{sec:FreqFns}, which then allows us to easily compare with 
the PN amplitude, which is always expressed as a function of frequency. Note that for this 
comparison we perform a frequency fit to our data during only the inspiral, which allows us to 
much more accurately capture the amplitude evolution; it is now much more necessary than in 
Sec.~\ref{sec:FreqFns} to have a reliable \emph{physical} fit.

\begin{figure}[t]
\centering
\includegraphics[width=80mm]{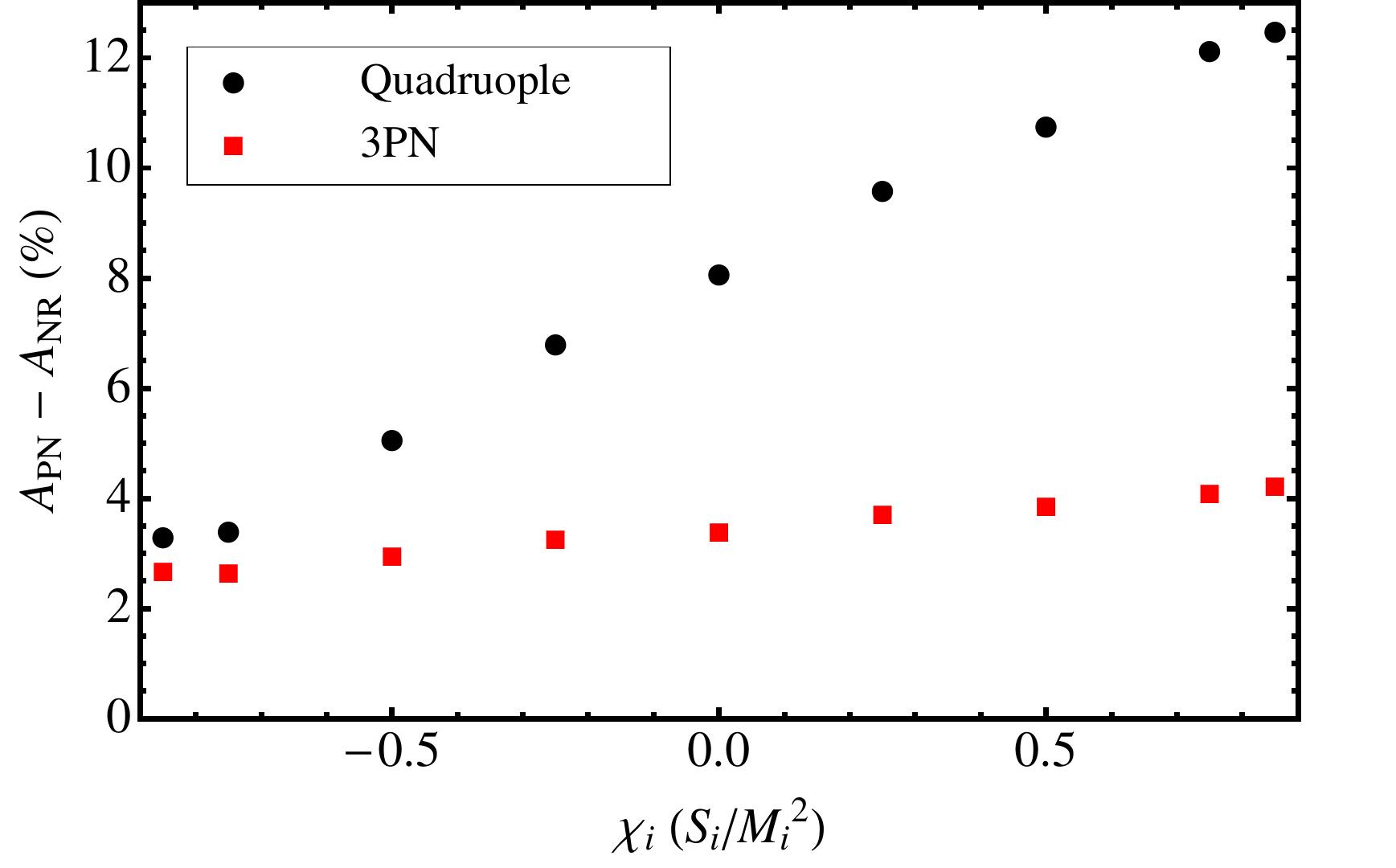}
\caption{
Average amplitude disagreement between PN and NR results, over the last
ten cycles up to $M\omega = 0.1$. The quadrupole PN amplitude error is only 
about 3\% for large anti-aligned spins, but rises to around 13\% for large aligned
spins. When the amplitude corrections are included up to 3PN-order, the PN 
amplitude error is only 3-4\% for all spin values.
}
\label{fig:PNAmpComparison}
\end{figure}

Fig.~\ref{fig:PNAmpComparison} shows the average disagreement between the PN and NR
amplitudes over the 10 cycles up to $M\omega = 0.1$, for the equal-mass spinning cases. 
The results using both the quadrupole and 3PN order amplitudes are shown. 
As seen in~\cite{Hannam:2007wf} the quadrupole amplitude disagreement rises to just over 
12\% for the highly spinning cases. The increase in disagreement is approximately linear with 
respect to the spin, and we predict that the maximum disagreement for extreme-spin black holes 
would be around 14\%. For large anti-parallel spins, the quadrupole amplitude performs much better, 
and drops to around 3\% for $\chi_i = -0.85$. 

When PN amplitude contributions up to 3PN (non-spinning)/2PN (spinning) order are used, the 
agreement with NR results is much better. In the nonspinning case it is 3\%, consistent with the results 
in~\cite{Boyle:2007ft}. (Note that the uncertainty in the extrapolated NR amplitude is around 1\%.) The 
variation with spin is small, rising to only 4\% in the high-spin hang-up cases, and falling to 
2.5\% in the high-spin anti-hang-up cases. We find similar results for the unequal-mass
cases, where the average disagreement is around 3\%. 

\section*{Discussion} 

We have presented the results of two sets of numerical simulations of black-hole binaries,
equal-mass binaries with equal, non-precessing spins with $\chi_i = S_i/M_i^2 \in [-0.85,0.85]$, 
and nonspinning unequal-mass binaries with $q = M_2/M_1 \in [1,4]$. These simulations
cover between six and 10 orbits before merger. The most accurate simulations have a 
numerical phase uncertainty during inspiral of 0.05\,rad, and a total accumulated phase error of
about 1.0\,rad. The phase uncertainties in the least accurate case
are 0.1\,rad during inspiral, and a total accumulated phase error of up to 15\,rad.
We have shown, however, that the uncertainty
estimates depend strongly on the alignment of the waveforms, and whether the results
are represented as functions of time or of GW frequency. The accuracy of the amplitude
of the $(\ell=2,m=2)$ mode of $\Psi_4$ is in general better than 1\% during inspiral, and 
between 2\% and 5\% during merger. 

For purposes of GW detection, the important quantity to consider is the mismatch-error
in the waveform. This is dominated by the errors due to the extraction of the GW signal
at a finite radius from the source. However, in all cases the mismatch error (minimized over
only time and phase) is below $10^{-5}$, meaning that the numerical waveforms are well
within the accuracy requirements for detection with current and planned ground-based 
detectors. 

These statements of waveform accuracy for detection apply only to the dominant mode and, 
more importantly, are only relevant when we consider binary masses such that the 
entire numerical waveform is within the sensitivity band of the detector, $M \apprge 120M_\odot$. 
For lower masses, longer waveforms are required, and in general can be produced by 
connecting post-Newtonian (PN) and numerical-relativity (NR) 
waveforms~\cite{Pan:2007nw,Ajith:2007xh,Santamaria:2010yb,Ajith:2007kx,
Boyle:2009dg}. The 
accuracy estimates given in this paper tell us \emph{nothing} about the accuracy of such
longer ``hybrid'' waveforms, because we cannot properly quantify the accuracy of the 
PN approximants.
We defer the discussion of the accuracy of hybrid 
waveforms, and the implications for the necessary length of numerical waveforms, 
to separate work~\cite{Hannam:2010ky}.

For now we consider the fidelity of PN results to full general relativity only in the regime 
where we also have NR results, i.e., in the last orbits before merger. We compare the 
PN and NR phase disagreement over the last 8-10 GW cycles before $M\omega = 0.1$
for two classes of PN approximant, TaylorT1 and TaylorT4. For nonspinning cases we
find that the performance of both approximants does not change drastically as the mass
ratio is increased to $q=4$, and this means that the TaylorT4 approximant continues to 
provide the best agreement, with an accumulated phase disagreement in the $q=4$ case
of 0.2\,rad, or 0.1\,rad if we consider the root-mean-square average of the phase
disagreement; see Fig.~\ref{fig:PNUMPhaseComp}. For spinning binaries, the 
two approximants include spin terms up to only 2.5PN order. The TaylorT1 approximant
nonetheless is fairly robust, while TaylorT4-truncated performs poorly for large spins, 
in particular large spins anti-aligned with the binary's orbital angular momentum.
The full TaylorT4 approximant performs well for all spins larger than $\chi_i = -0.75$.

Finally, we study the accuracy of the PN wave amplitude, and find that when the highest
order amplitude corrections are included (3PN for nonspinning binaries, and 2PN for spinning
cases), the amplitude error is no more than 4\%. This is in contrast to the quadrupole
amplitude,
which can over-estimate the true physical amplitude by up to 13\% at black hole dimensionless
spins of $\chi_i = +0.85$ corresponding to an increase of 44\% in detection rates. Note that
precisely the cases with the largest SNR (spins aligned with the angular momentum) are also 
those with the largest PN amplitude errors.

\section*{Acknowledgments}

We thank Mike Boyle for extensive cross-checks of the spin-dependent terms in the phase
and amplitude of the PN approximants used in this paper.

M. Hannam was supported by an FWF Lise-Meitner Fellowship (M1178-N16)
and a Science and Technology Facilities Council Advanced Fellowship (ST/H008438/1),
and thanks the Department of Physics at the University of the Balearic Islands 
for hospitality while some of this work was carried out.
S. Husa was supported by
grant FPA-2007-60220 from the Spanish Ministry of Science,
the Spanish MICINN’s Consolider-Ingenio 2010 Programme under grant 
MultiDark CSD2009-00064, and DAAD grant D/07/13385.

This work was supported in part by DFG grant SFB/Transregio~7
``Gravitational Wave Astronomy'' and the DLR (Deutsches Zentrum f\"ur Luft-
und Raumfahrt). D. M\"uller was additionally supported by the DFG Research
Training Group 1523 ``Quantum and Gravitational Fields''.

{\tt BAM} simulations were carried out at LRZ Munich, ICHEC Dublin,
on the Vienna Scientific Cluster (VSC), 
at MareNostrum at Barcelona Supercomputing Center --
Centro Nacional de Supercomputaci\'on (Spanish National
Supercomputing Center), and CESGA, Santiago the Compostela.

\appendix

\section{Apparent-horizon and puncture estimates of the black-hole masses} 
\label{app:masses}

There are two methods that are commonly used to estimate the masses
of black holes in puncture data, in addition to analyzing apparent horizons. The first, which is generally applicable to all
black-hole data, is to make use of the area of the apparent horizon, $A$. 
The black hole's ``irreducible mass'' $M_{irr}$ is given by $M_{irr} = \sqrt{A/16\pi}$,
and the total mass can be estimated by~\cite{Christodoulou70} \begin{equation}
M^2 = M_{irr}^2 + \frac{S^2}{4 M_{irr}^2}.
\end{equation} 

A second method is to make use of the asymptotic properties of the wormhole 
puncture data. Each puncture represents an extra asymptotically flat end of the 
slice, and the ADM mass calculated at each ``extra'' end can be considered as a 
measure of the mass of that black hole. In the puncture-data construction, the
momentum constraint is solved analytically by the Bowen-York conformal extrinsic
curvature, and the Hamiltonian constraint is solved numerically to give the function
$u$ in the ansatz~\cite{Brandt:1997tf}, \begin{equation}
\psi = 1 + \frac{m_1}{2r_1} + \frac{m_2}{2r_2} + u,
\end{equation} where $m_i$ parametrizes the mass of the $i$th black hole, and $r_i$
is the coordinate distance to the $i$th black hole. The resulting data represent two 
black holes on a three-sheeted topology. One sheet contains two black holes, and
represents the physical space that we want to describe. Each black hole has an 
extra sheet associated with it, which extends to an extra asymptotically flat end, and 
in the puncture construction those ends are compactified to points, or ``punctures''. 

To calculate this mass, we require only 
the value of the function $u$ at the puncture. The mass is then given by
\begin{equation}
M_i = m_i \left(1 + u_i + \frac{m_j}{2 D} \right),
\end{equation} where $D$ is the coordinate distance between the 
two punctures. A derivation of this expression is given in~\cite{Brandt:1997tf}. 
The two measures of the mass that we have discussed are shown to agree 
within numerical uncertainty in the case of nonspinning black holes 
in~\cite{Tichy:2003qi}. Since the ADM mass at the puncture can be easily 
calculated directly from the initial data with high precision, it has become a 
standard tool in assessing the mass of black holes in puncture data.

However, as discussed in~\cite{Hannam:2009ib}, this is only a reasonable measure
of the black-hole mass for nonspinning black holes. A heuristic explanation for 
this effect is that the fall-off of the extrinsic curvature for a boosted Bowen-York black 
hole is far faster towards the
extra asymptotically flat ends as it is towards the ``physical'' end, and so the extra
sheets of the topology contain far less junk radiation than the physical sheet, and the ADM mass of
each of those sheets is not contaminated by very much junk radiation. In the spinning case, however,
the fall-off on the second sheet is the same as on the physical sheet, and so the extra
sheets each contain roughly the same junk radiation as the physical space, and only for
low spins will the ADM mass at the puncture be a good measure of the black-hole mass.

As an illustration of this effect, the values of the black-hole mass as given by the 
two methods are shown in Table~\ref{tab:masses}. 
\begin{table}[htbp]
\caption{\label{tab:masses}
The uncertainty in the apparent-horizon mass is about 0.01\%, and so the
horizon and puncture masses agree within uncertainty for the
$S_i/M_i^2 = \{0,0.25\}$ cases. For higher masses, however, the
discrepancy between the horizon and puncture masses is clear.
}
\begin{tabular}{||c|c|c|c|}
\hline
$S_i/M_i^2$  &  $M_{AH}$  &    $M_{puncture}$  & Error (\%) \\
\hline
0           &  0.50001   &  0.50000  & 0.002  \\
0.25     &  0.49998   &  0.50000  & 0.004  \\
0.5       &  0.49977   & 0.50000   & 0.046   \\
0.75     &  0.49815  &  0.50000   & 0.370  \\
0.85     &  0.49577  & 0.50000    & 0.846  \\
\hline
\end{tabular}
\end{table}

For the simulations presented in this paper, the results were first produced using
the puncture-mass estimates. They were then rescaled according to the results in
Tab.~\ref{tab:masses}. A rescaling of mass will have an overall effect on the time-scale
of the simulations, but we found that even in the highest spin case the effect was 
negligible. This is most important in the comparison with PN approximants in
Sec.~\ref{sec:pncomparison}, where the PN and NR results are compared assuming 
the same mass scale. But we find that the phase disagreement between the NR and
NR results is much larger than the error due to using the incorrect black-hole mass,
and does not noticeably alter the results in, for example, Fig.~\ref{fig:PNSpinPhaseComp}.

\bibliography{NonPrecessing}

\end{document}